\documentclass[showpacs,preprintnumbers,amsmath,amssymb,twocolumn]{revtex4}

\usepackage{graphicx}
\usepackage{dcolumn}
\usepackage{bm}

\begin{document}

\title{Neutral Color Superconductivity Including Inhomogeneous Phases\\
       at Finite Temperature}
\author{\normalsize{Lianyi He\footnote{Email address: hely04@mails.tsinghua.edu.cn}$^1$, Meng Jin\footnote{Email address: jin-m@mail.tsinghua.edu.cn
        }$^{1,2}$ and Pengfei Zhuang\footnote{Email address:
        zhuangpf@mail.tsinghua.edu.cn}$^1$}}
\affiliation{1 Physics Department, Tsinghua University, Beijing
100084, China\\
2 Institute of Particle Physics, Central China Normal University,
Wuhan 430070, China}

\begin{abstract}
We investigate neutral quark matter with homogeneous and
inhomogeneous color condensates at finite temperature in the frame
of an extended NJL model. By calculating the Meissner masses
squared and gap susceptibility, the uniform color superconductor
is stable only in a temperature window close to the critical
temperature and becomes unstable against LOFF phase, mixed phase
and gluonic phase at low temperatures. The introduction of the
inhomogeneous phases leads to disappearance of the strange
intermediate temperature 2SC/g2SC and changes the phase diagram of
neutral dense quark matter significantly.
\end{abstract}

\date{\today}

\maketitle

\section {Introduction}
The emergence of gapless color
superconductivity\cite{huang,shovkovy,huang2,alford} promoted
great interest in the study of dense quark
matter\cite{itoh,witten}. A crucial problem is whether the two
flavor gapless (g2SC) or breached pairing\cite{BP} color
superconductivity is stable. The first instability is the
thermodynamical instability  when the chemical potential
difference between the two species is fixed. This is called Sarma
instability\cite{sarma}. It is now widely accepted that the Sarma
instability can be cured under charge neutrality constraint due to
the long range electromagnetic gauge
interaction\cite{shovkovy,huang2}. The second instability is the
chromomagnetic instability described by negative Meissner mass
squared\cite{huang3,casalbuoni,alford2,fukushima,he} or negative
superfluid density\cite{wu,he2}. In this case, some inhomogeneous
condensed phases are energetically
favored\cite{giannakis,giannakis2,giannakis3,huang4,hong,gorbar,gorbar2,he3}.
A specific character of the chromomagnetic instability is the
negative Meissner masses squared for the 4-7th gluons in the
gapped 2SC phase\cite{he5}. The third instability is the
instability against the phase separation or mixed phase when the
surface energy and long range gauge interaction are excluded.

All these instabilities are confirmed only at zero temperature and
in the weak coupling region. While it is recently argued that the
breached pairing phase will be free from negative superfluid
density or Meissner mass squared in the strong coupling BEC
region\cite{pao,smit,kita,he4}, the diquark condensate at moderate
baryon density seems not in this region. As for the temperature
effect, in the study of isospin asymmetric nuclear
matter\cite{sedrakian,Isayev}, two flavor color-superconducting
quark matter\cite{huang2} and general breached pairing
superfluidity\cite{liao}, the fermion pairing correlation
behaviors strangely: The superfluid/supercondutor order parameter
is not a monotonously decreasing function of temperature and its
maximum is located at finite temperature. Especially, for large
enough density asymmetry between the two species at fixed coupling
or for weak enough coupling at fixed density asymmetry, the
superfluidity appears at intermediate temperature but disappears
at low temperature. This strange temperature behavior is quite
universal and also discussed in the study of atomic Fermi gas with
density imbalance\cite{chen,sedrakian2,he6}. It is
found\cite{chen,he6} that the superfluid density and number
susceptibility of the uniform superfluid phase are positive only
in a temperature window near the critical temperature. From the
recent instability analysis at finite temperature\cite{he6}, if
LOFF phase is taken into account, this strange temperature
behavior will be washed out and the superfluid order parameter
remains a monotonous function of temperature. Therefore, for the
color superconductivity at moderate density, we may have the
following estimations: (1) The superconductor is free from the
chromomagnetic instability and also stable against the mixed phase
at temperatures near the critical value. (2) The phase diagram of
neutral dense quark matter is significantly changed if some
non-uniform phase such as LOFF is taken into account. In this
paper we will show that these two estimations are true.

The paper is organized as follows. In Section \ref{s2} we study a
naive two-species model which possesses the basic mechanism for
positive Meissner mass squared and gap susceptibility at finite
temperature. We then calculate in an extended NJL model the
Meissner masses squared for gluons and the gap susceptibility in
two flavor color-superconducting quark matter with charge
neutrality in Section \ref{s3} and investigate the LOFF phase and
gluonic phase with charge neutrality in Section \ref{s4}. We
summarize in Section \ref{s5}.

\section {Results in a Toy Model}
\label{s2}
In this section we study a toy model containing two species of
fermions proposed in\cite{alford2,he,kita}. While this model only
reveals the chromomagnetic instability for the 8th gluon and the
important non-abelian type chromomagnetic instability is excluded,
it can help us to better understand why the choromomagnetic
instability can be cured at finite temperature. The Lagrangian of
the model is defined as
\begin{equation}
{\cal L}=\bar{\psi}\left(i\gamma^\mu\partial_\mu-m\right)\psi
+G\left(\bar{\psi}^C i\gamma_5\tau_1\psi\right)\left(\bar{\psi}
i\gamma_5\tau_1\psi^C\right)\ ,
\end{equation}
where $\psi=(\psi_1,\psi_2)^T$ and
$\bar{\psi}=(\bar{\psi}_1,\bar{\psi}_2)$ are Dirac spinors
including two species of fermions, $\psi^C=C\bar\psi^T$ and
$\bar\psi^C=\psi^T C$ are charge-conjugate spinors,
$C=i\gamma^2\gamma^0$ is the charge conjugation matrix, $m$ is the
fermion mass, $G$ is the attractive coupling, and $\tau_1$ is the
first Pauli matrix in the two-species space. In the following we
take $m=0$ which corresponds to the quark matter at high baryon
density. It is quite convenient to work with the Nambu-Gorkov
spinors defined as $\Psi=(\psi, \psi^C)^T$ and
$\bar{\Psi}=(\bar{\psi}, \bar{\psi}^C)$. We assume that the
attractive coupling is not very strong and the BCS mean field
theory remains a good treatment at finite temperature. The study
on possible BCS-BEC crossover in relativistic fermion superfluid
can be seen in \cite{RBEC1,RBEC2}. In the mean field
approximation, the pair fields are replaced by their expectation
values which serve as order parameters of superfluidity.
Introducing the order parameters
\begin{equation}
\Delta=-2G\left\langle\bar{\psi}^C
i\gamma_5\tau_1\psi\right\rangle\ , \ \ \
\Delta^*=-2G\left\langle\bar{\psi}
i\gamma_5\tau_1\psi^C\right\rangle\
\end{equation}
and taking them to be real, we can express the thermodynamic
potential as
\begin{equation}
\Omega ={\Delta^2\over 4G}-{T\over 2}\sum_n\int{d^3 \vec{p}\over
(2\pi)^3}\textrm{Tr} \ln {\cal S}^{-1}(i\omega_n,\vec{p})
\end{equation}
with the mean field fermion propagator in the 4-dimensional
Nambu-Gorkov$\otimes$two-species space,
\begin{equation}
\label{g2}
{\cal
S}^{-1}(i\omega_n,\vec{p})=\left(\begin{array}{cc}
[G_0^+]^{-1}&i\gamma_5\Delta\tau_1
\\ i\gamma_5\Delta\tau_1&[G_0^-]^{-1}\end{array}\right),
\end{equation}
where $[G_0^\pm]^{-1}=(i\omega_n\pm\hat{\mu})\gamma_0-
\vec{\gamma}\cdot\vec{p}$ are the inverse of free fermion
propagators with frequency $\omega_n=(2n+1)\pi T,\ n=1,2,...$. We
have introduced the chemical potentials $\mu_1$ and $\mu_2$ for
the two species via adding the term ${\cal
L}_\mu=\bar{\psi}\hat{\mu}\gamma^0\psi$ to the original Lagrangian
with $\hat{\mu}=diag(\mu_1,\mu_2)$. After a straightforward
calculation, the thermodynamic potential can be evaluated as
\begin{equation}
\Omega ={\Delta^2\over 4G}-\sum_{I=1}^4\int{d^3 {\vec p}\over
(2\pi)^3}W(E_I)
\end{equation}
with $W(E_I)=E_I+2T\ln\left(1+e^{-E_I/T}\right)$, where the sum
runs over all fermionic quasiparticles. The quasiparticle
dispersions $E_I(p)$ calculated via $\det{\cal S}^{-1}(p)=0$ are
given by
\begin{eqnarray}
\label{e1234}
&& E_1(p)=E_\Delta^- -\delta\mu,\ \ \
E_2(p)=E_\Delta^-
+\delta\mu,\nonumber\\
&& E_3(p)=E_\Delta^+ -\delta\mu,\ \ \ E_4(p)=E_\Delta^+
+\delta\mu,
\end{eqnarray}
where we have defined the notation
$E_\Delta^\pm=\sqrt{(p\pm\bar{\mu})^2+\Delta^2}$ and introduced
the average chemical potential $\bar\mu=(\mu_1+\mu_2)/2$ and the
chemical potential mismatch $\delta\mu=(\mu_2-\mu_2)/2$. The gap
equation for $\Delta$ can be derived via minimizing the
thermodynamic potential,
\begin{equation}
\label{gap}
\partial\Omega/\partial\Delta=0
\end{equation}
and the fermion numbers are determined by the definition,
\begin{equation}
\label{number} n_1=-\partial \Omega/\partial \mu_1,\ \ \
n_2=-\partial \Omega/ \partial \mu_2.
\end{equation}
The order parameter $\Delta$ under constraint of fixed numbers
$n_1$ and $n_2$ can be calculated by solving the gap equation
(\ref{gap}) together with the number equations (\ref{number}).
Since the gap equation for $\Delta$ suffers ultraviolet
divergence, we introduce a three momentum cutoff $\Lambda$ to
regularize it.

\begin{figure}[!htb]
\begin{center}
\includegraphics[width=6cm]{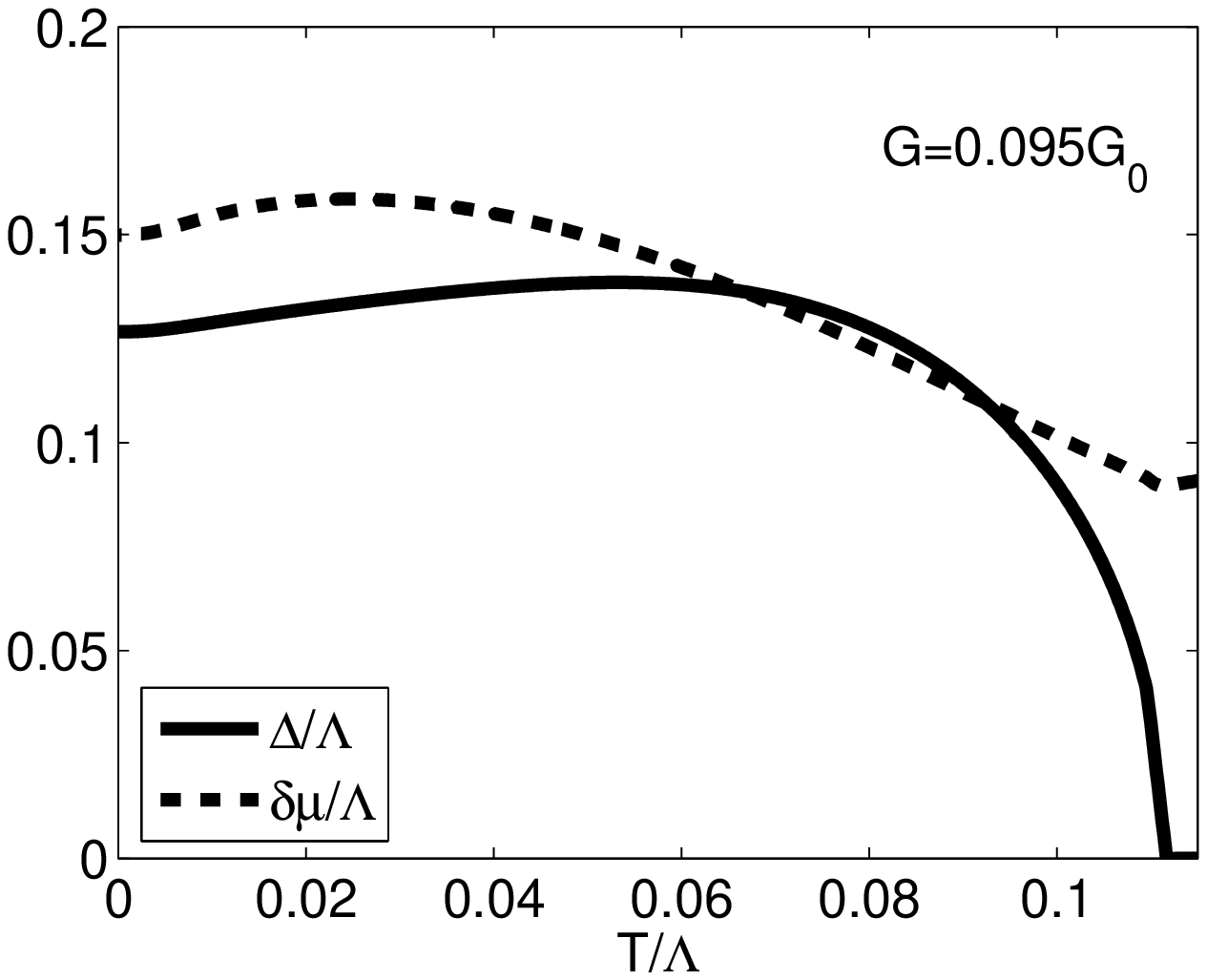}
\includegraphics[width=6cm]{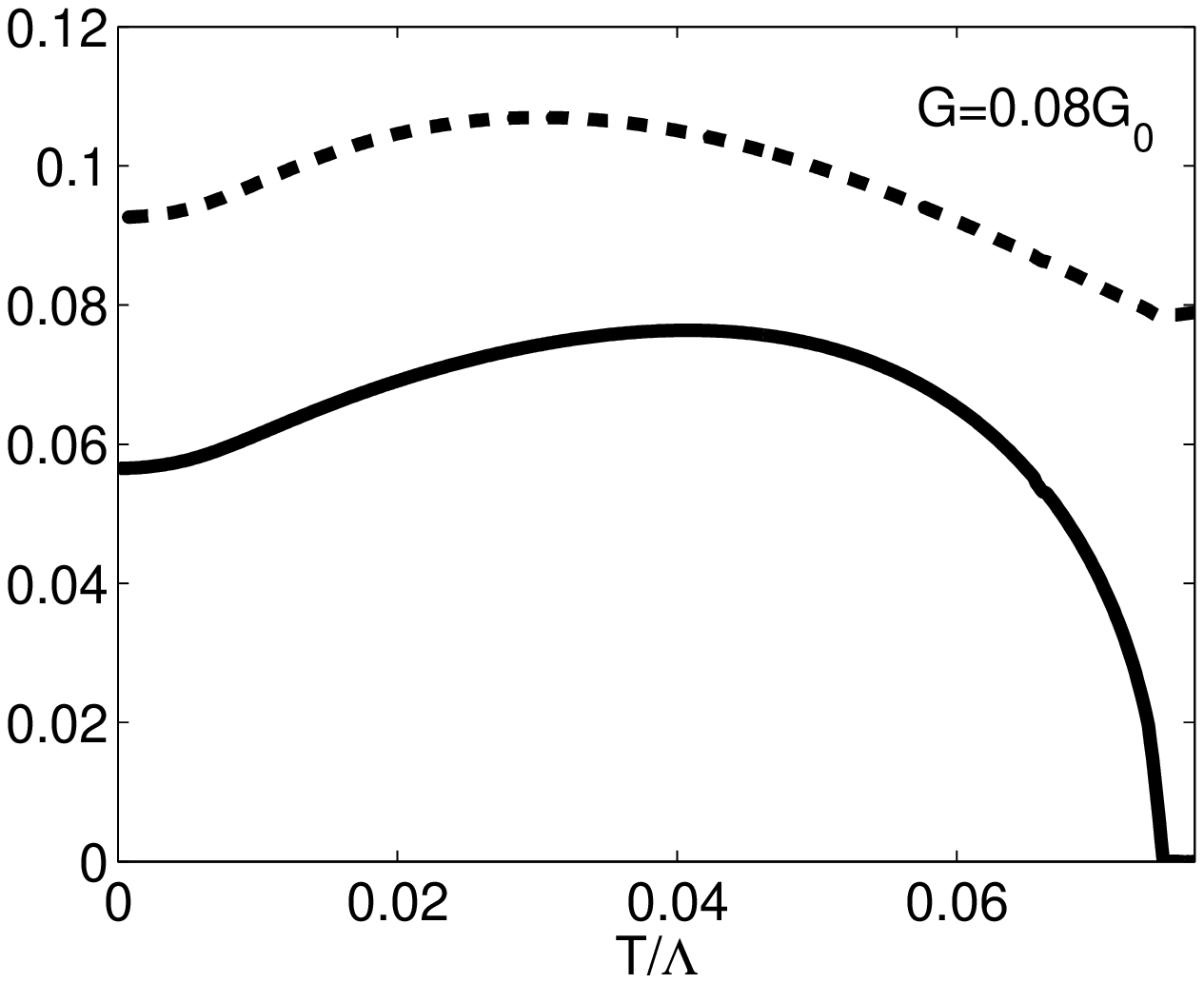}
\includegraphics[width=6cm]{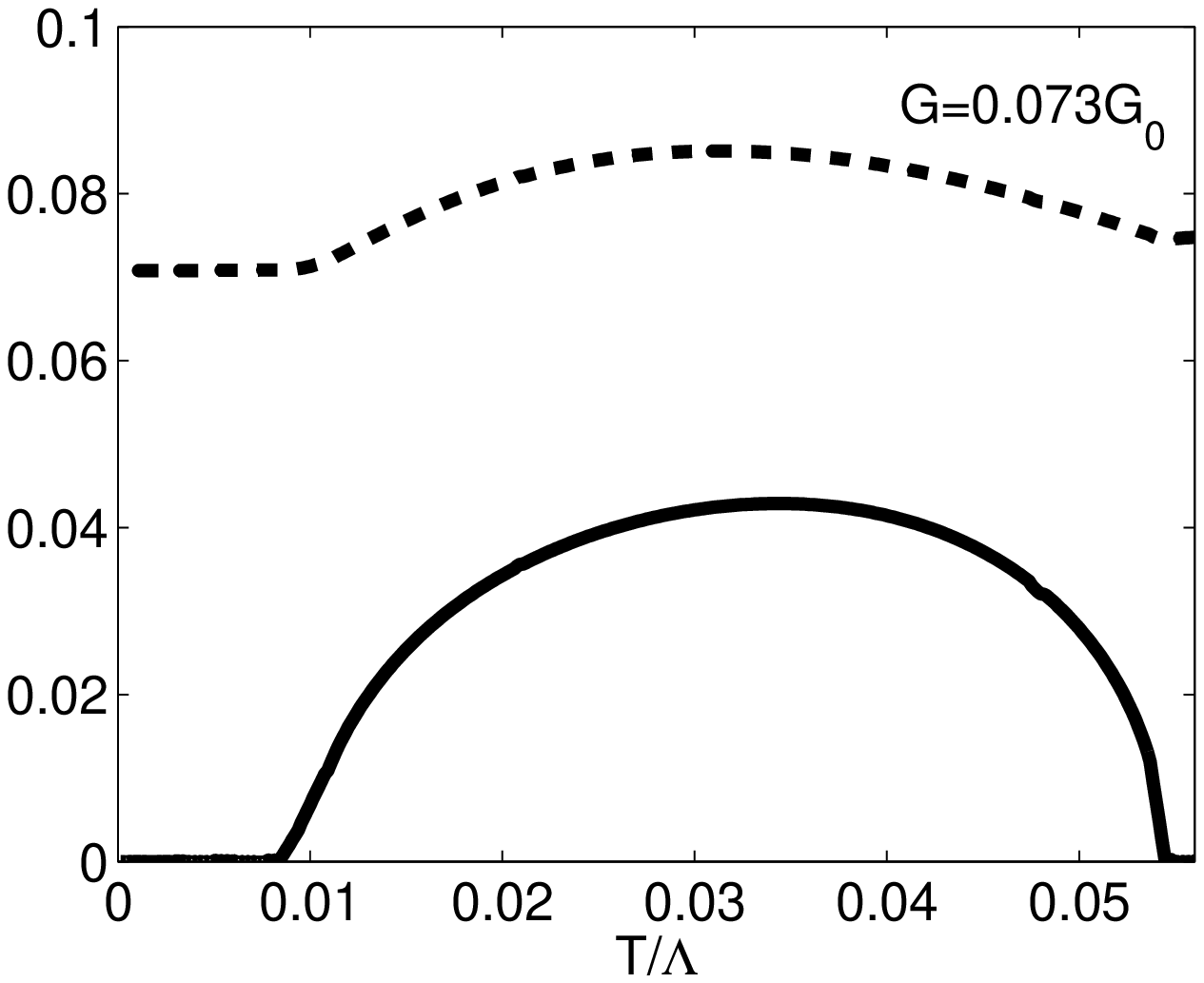}
\caption{The pairing gap $\Delta$ and chemical potential mismatch
$\delta\mu$ as functions of temperature $T$ for three coupling
values. $\Delta, \delta\mu$ and $T$ are all scaled by the cutoff
$\Lambda$. \label{fig1}}
\end{center}
\end{figure}

To model the two flavor dense quark matter with beta equilibrium
and charge neutrality, we choose the coupling constant $G\leq
0.1G_0$ with $G_0=4\pi^2/\Lambda^2$ to ensure weak coupling and
fix the averaged chemical potential $\bar{\mu}=0.615\Lambda$ and
the number ratio $n_1/n_2 =2$. In Fig.\ref{fig1} we show the
temperature behavior of $\Delta$ and $\delta\mu$. For the coupling
$G=0.095G_0$, the gapless state satisfying the condition
$\delta\mu > \Delta$ appears at low and high temperature regions
but disappear at intermediate temperature. For weaker couplings
$G=0.08G_0$ and $0.073G_0$, the superfluid is always gapless. For
weak enough coupling such as $G=0.073G_0$, the superfluidity
appears only at intermediate temperature.

For a symmetric system with $n_1=n_2$, it is well-known that the
temperature effect destroys the pairing, and the pairing gap is a
monotonously decreasing function of temperature. However, for an
asymmetric system with $n_1\neq n_2$, the two species of fermions
have mismatched Fermi surfaces, and the temperature effect not
only deforms and reduces the Fermi surfaces which melts the gap,
but also makes the overlap region of the two species wider which
favors the pairing. The competition of the two opposite effects
results in a non-monotonous temperature behavior of the pairing
gap. This is the reason why in the low temperature region in
Fig.\ref{fig1} the gap increases as the temperature increases and
even disappears when the coupling is small enough.

\subsection {Meissner Mass}
Let us assume that the fermions couple to a $U(1)$ gauge field
$A_\mu$ with coupling $g$, the superfluid becomes a superconductor
where the gauge field obtain the so called Meissner mass. The
Meissner mass can be calculated from the polarization tensor
\begin{equation}
\Pi^{\mu\nu}(k)=\frac{T}{2}\sum_n\int\frac{d^3\vec{p}}{(2\pi)^3}\textrm{Tr}
\left[\hat{\Gamma}^\mu{\cal S}(p)\hat{\Gamma}^\nu{\cal
S}(p-k)\right]
\end{equation}
with the vertex
$\hat{\Gamma}^\mu=g\gamma^\mu\sigma_3\otimes\tau_0$, where
$\sigma_3$ is the third Pauli matrix in the two-species space,
$\tau_0$ is the identity matrix in the Nambu-Gorkov space, and the
propagator ${\cal S}$ can be obtained from its inverse (\ref{g2})
by taking the technics used in \cite{he2}. The Meissner mass $m_A$
is defined as
\begin{equation}
m_A^2=\frac{1}{2}\lim_{\vec{p}\rightarrow0}(\delta_{ij}-k_i
k_j/|{\bf k}|^2)\Pi^{ij}(k_0=0,\vec{k}).
\end{equation}

The calculation of the Meissner mass is quite straightforward but
tedious, we quote here only the final result,
\begin{equation}
m_A^2=m_{d}^2+m_{p}^2,
\end{equation}
the diamagnetic term $m_d^2$ and paramagnetic term $m_p^2$ are
explicitly expressed as
\begin{eqnarray}
m_{d}^2&=&\frac{4g^2}{3}\int
\frac{d^3\vec{p}}{(2\pi)^3}\big[\chi^-(p)\left(f(E_1)+f(E_2)-1\right)\nonumber\\
&&\ \ \ \ \ \ \ \ \ \ \ \ \ \ \ \ +\chi^+(p)\left(f(E_3)+f(E_4)-1\right)\big]+m_\Lambda^2\nonumber\\
m_{p}^2&=&\frac{g^2}{3}\int
\frac{d^3\vec{p}}{(2\pi)^3}\sum_{I=1}^4f^\prime(E_I),
\end{eqnarray}
where $f(x)$ is the Fermi-Dirac distribution function,
$f^\prime(x)$ is its first order derivative, and the functions
$\chi^\pm(p)$ are defined as
\begin{equation}
\chi^\pm(p)=\pm\frac{(E_\Delta^\pm)^2-E_0^+E_0^-+\Delta^2}{E_\Delta^\pm\left((E_\Delta^+)^2-(E_\Delta^-)^2\right)}
\simeq\frac{1}{2p}\frac{E_0^\pm}{E_\Delta^\pm}
\end{equation}
with $E_0^\pm = p\pm\bar\mu$. Since the diamagnetic term suffers
ultraviolet divergence, it contains a subtraction term
\begin{equation}
m_\Lambda^2=\frac{4g^2}{3}\int
\frac{d^3\vec{p}}{(2\pi)^3}\frac{1}{p}=\frac{g^2\Lambda^2}{3\pi^2}
\end{equation}
to ensure $m_A^2(\Delta=0)=0$.

Above the critical temperature $T_c$ where the pairing gap
$\Delta$ vanishes, we have $\chi^\pm(p)=1/(2p)$ and
\begin{eqnarray}
m_{d}^2&=&\frac{2g^2}{3}\int
\frac{d^3\vec{p}}{(2\pi)^3}\frac{1}{p}\sum_{i=1}^2\bigg[f(p-\mu_i)+f(p+\mu_i)\bigg],\nonumber\\
m_{p}^2&=&\frac{g^2}{3}\int
\frac{d^3\vec{p}}{(2\pi)^3}\sum_{i=1}^2\bigg[f^\prime(p-\mu_i)+f^\prime(p+\mu_i)\bigg].
\end{eqnarray}
Taking partial integration we find
\begin{eqnarray}
\frac{1}{3}\int
\frac{d^3\vec{p}}{(2\pi)^3}f^\prime(p\pm\mu)=-\frac{2}{3}\int
\frac{d^3\vec{p}}{(2\pi)^3}\frac{1}{p}f(p\pm\mu)
\end{eqnarray}
for any $T$ and $\mu$. Therefore, the total Meissner mass $m_A^2$
is zero in the normal phase, as it is expected.

In weak coupling with $\Delta\ll\bar{\mu}$ and
$\delta\mu\ll\bar{\mu}$, the Meissner mass squared at zero
temperature can be well approximated as\cite{huang3}
\begin{equation}
m_A^2\approx\frac{g^2\bar{\mu}^2}{3\pi^2}\left[1-\frac{\delta\mu\theta(\delta\mu-\Delta)}{\sqrt{\delta\mu^2-\Delta^2}}\right].
\end{equation}
In the gapless phase with $\Delta<\delta\mu$ the negative Meissner
mass squared is quite different from
$m_A^2=g^2\bar{\mu}^2/3\pi^2>0$ in the symmetric BCS case where
there is no paramagnetic contribution.

Now the question is: Is the Meissner mass squared always negative
below the critical temperature $T_c$? To answer this question, we
derive the critical behavior of the Meissner mass squared at
$T_c$. Since the gap equation for $\Delta$ below $T_c$ contains
only $\Delta^2$ and $\Delta$ continuously approaches to zero at
$T_c$, we have the critical behavior for the order parameter
\begin{equation}
\Delta(T) \propto\left(1-T/T_c\right)^{1/2},\ \ \ T\rightarrow
T_c.
\end{equation}
As a consequence, the Meissner mass squared behaviors as
\begin{equation}
m_A^2(T) \propto1-T/T_c>0,\ \ \ T\rightarrow T_c.
\end{equation}
Combining $m_A^2 < 0$ at zero temperature and $m_A^2 > 0$ for $T$
approaching to $T_c$, $m_A^2$ should be negative at low
temperature but positive in a temperature window close to $T_c$.

\begin{figure}[!htb]
\begin{center}
\includegraphics[width=6cm]{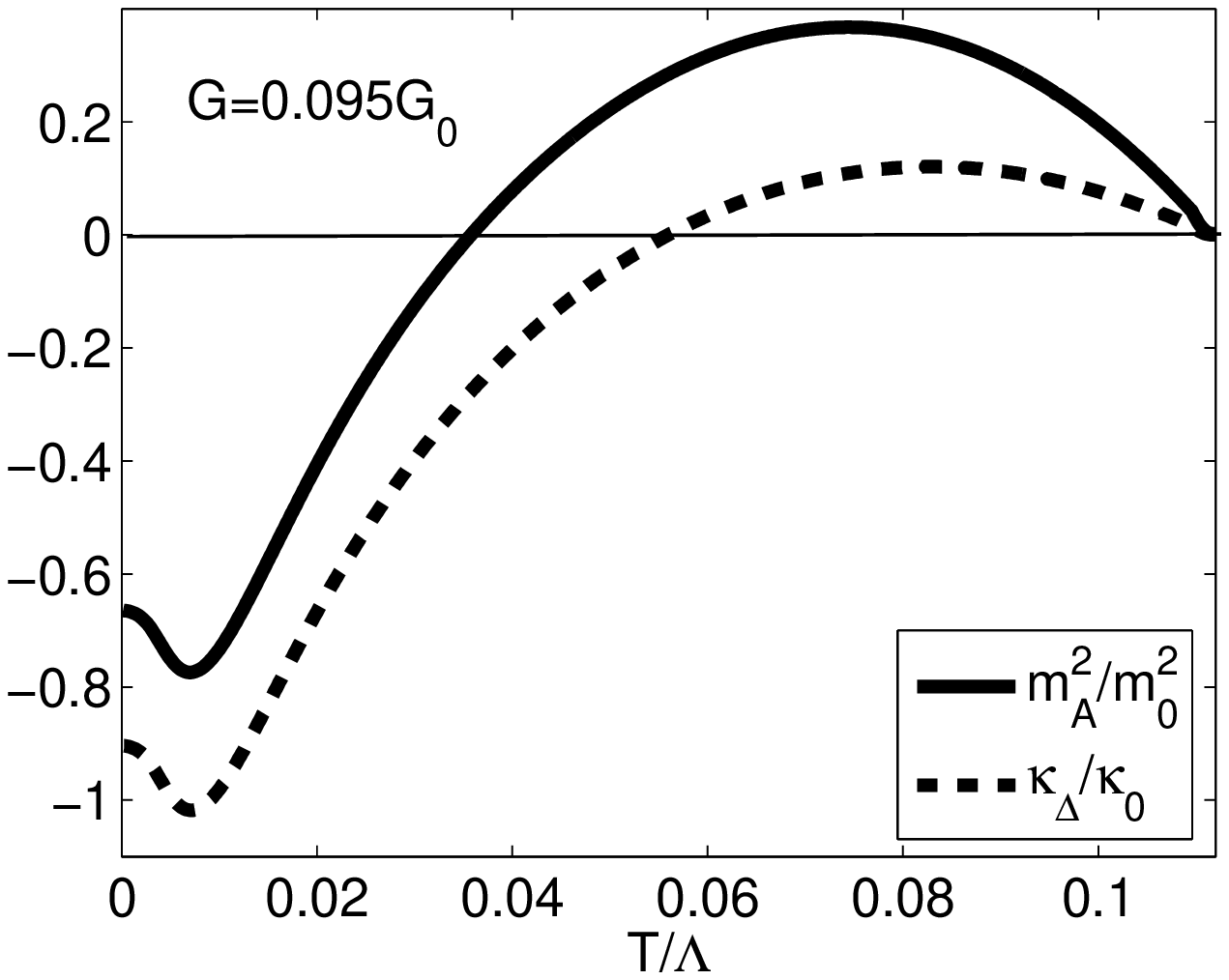}
\includegraphics[width=6cm]{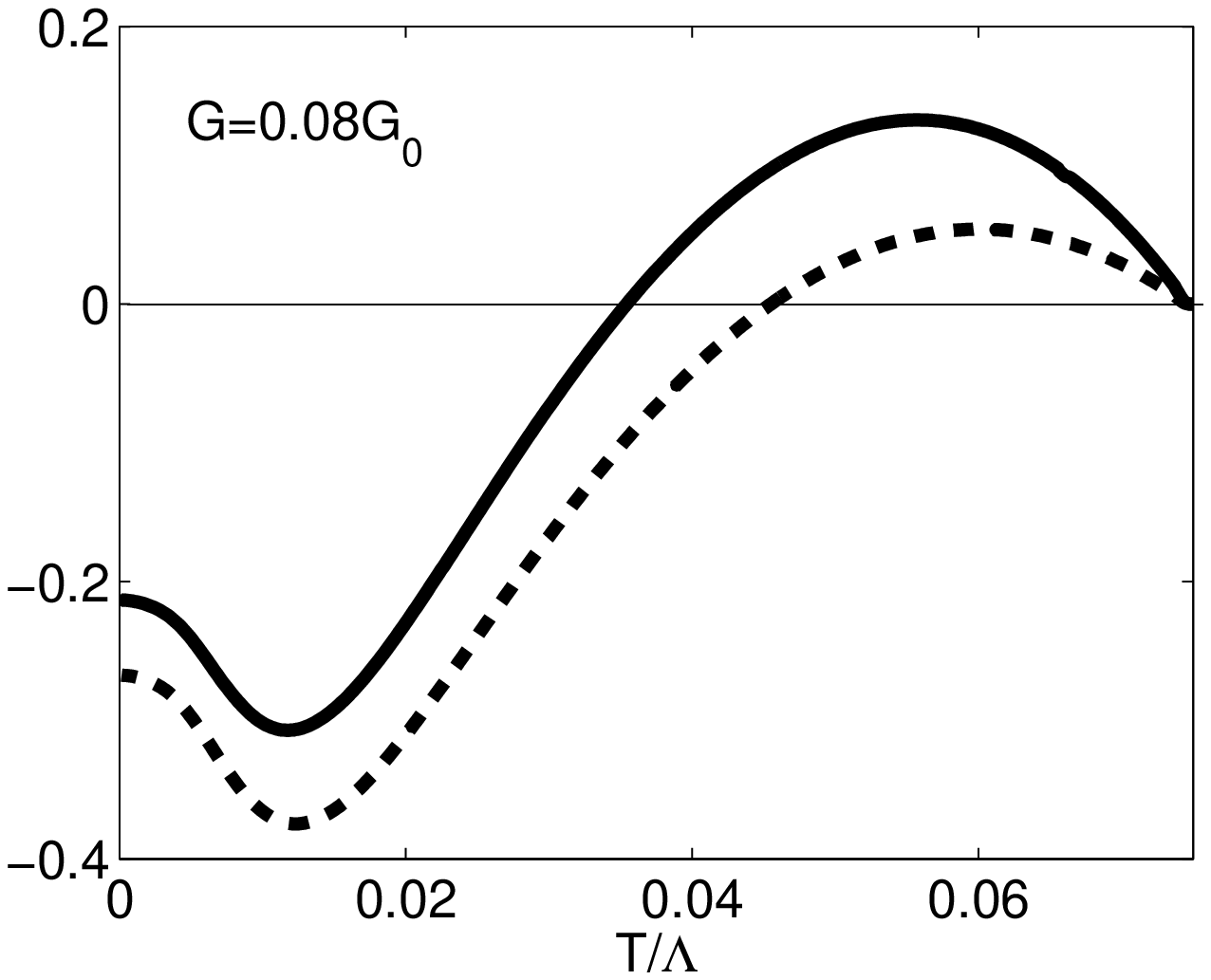}
\includegraphics[width=6cm]{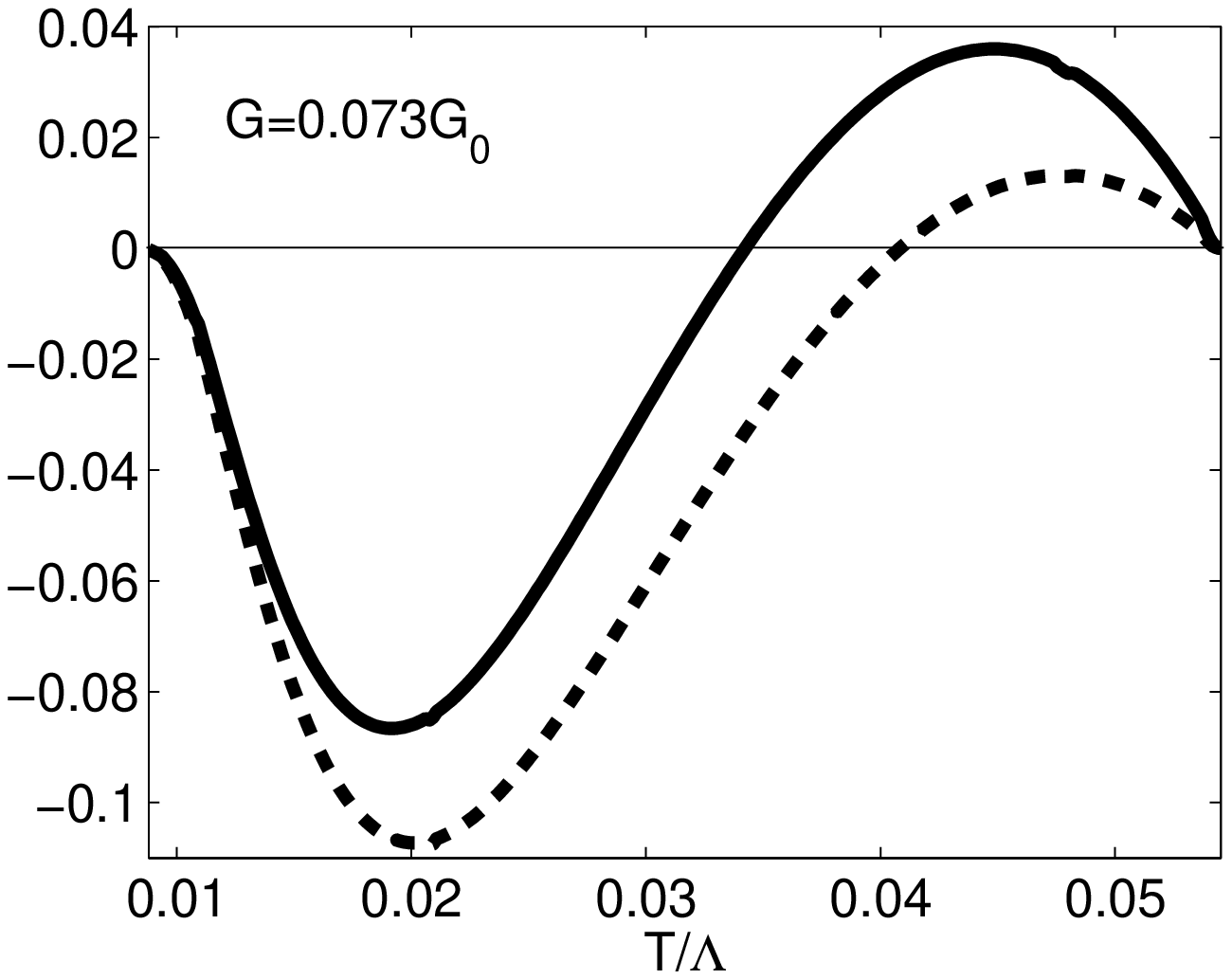}
\caption{The Meissner mass squared $m_A^2$ scaled by
$m_0^2=g^2\bar{\mu}^2/3\pi^2$ and the gap susceptibility
$\kappa_\Delta$ scaled by $\kappa_0=2\bar{\mu}^2/\pi^2$ as
functions of $T/\Lambda$ for three coupling values. \label{fig2}}
\end{center}
\end{figure}

In Fig.\ref{fig2} we show the temperature behavior of the Meissner
mass squared for three coupling values. In any case, there exists
a temperature window close to the critical temperature where the
Meissner mass squared $m_A^2$ is positive which means that the
gapless phase is magnetically stable and hence stable against the
LOFF phase.

\subsection {Density Fluctuations}
The general stability condition against changes in number
densities for a two-component system is described by the total
free energy of the system\cite{viverit,cohen}, $F=\int d^3{\bf
x}{\cal F}[n_\sigma({\bf x})] (\sigma=1,2)$. Considering its
fluctuations induced by small number changes $\delta n_\sigma({\bf
x})$, the first-order variation $\delta F$ vanishes automatically
due to the charge conservation, $\int d^3{\bf x}\delta
n_\sigma({\bf x})=0$, and the second-order variation $\delta^2F$
is expressed in the quadratic form
\begin{equation}
\delta^2F=\frac{1}{2}\int d^3{\bf
x}\sum_{\sigma,\sigma'=1,2}\frac{\partial^2{\cal F}}{\partial
n_\sigma\partial n_{\sigma'}}\delta n_\sigma\delta n_{\sigma'}.
\end{equation}
Therefore, to achieve a stable homogeneous phase, the $2\times 2$
matrix $\partial^2{\cal F}/\partial n_\sigma\partial n_{\sigma'}$
should be positively definite, namely, it has only positive eigen
values. From the relation between the free energy ${\cal
F}(n_\sigma)$ and thermodynamic potential $\Omega(\mu_\sigma)$,
$\Omega={\cal F}+\mu_1n_1+\mu_2n_2$, it is easy to check that the
stability condition to have positively definite matrix
$\partial^2{\cal F}/\partial n_\sigma\partial n_{\sigma'}$ is
equivalent to the condition to have positively definite number
susceptibility matrix $-\partial^2\Omega/\partial
\mu_\sigma\partial \mu_{\sigma'}$.

For systems without mass difference between the two species, the
condition to have positive eigenvalues of
$-\partial^2\Omega/\partial \mu_\sigma\partial \mu_{\sigma'}$ can
be reduced to the condition\cite{pao} that the imbalance number
susceptibility
$\chi=-(\partial^2\Omega/\partial\delta\mu^2)_\mu=\left(\partial\delta
n/\partial\delta\mu\right)_{\bar{\mu}}$ should be positive. For
$\chi<0$, the density difference $\delta n=n_1-n_2$ increases with
decreasing chemical potential difference $\delta\mu$, which is
certainly unphysical and means that the uniform phase is unstable
against density fluctuations. Employing the gap equation which
determines the condensate as a function of chemical potentials, we
can express the imbalance number susceptibility $\chi$ as a direct
and an indirect part,
\begin{equation}
\chi=\left(\frac{\partial\delta
n}{\partial\delta\mu}\right)_{\bar{\mu},\Delta}+\left(\frac{\partial\delta
n}{\partial\Delta}\right)_{\bar{\mu},\delta\mu}\left(\frac{\partial\Delta}{\partial\delta\mu}\right)_{\bar{\mu}}.
\end{equation}
From the expression
\begin{equation}
\label{deltan}
\delta n=\int{d^3{\vec p}\over
(2\pi)^3}\left[f(E_1)-f(E_2)+f(E_3)-f(E_4)\right]
\end{equation}
and the gap equation
$\left(\partial\Omega/\partial\Delta\right)_{\bar{\mu},\delta\mu}=0$,
we have\cite{he6}
\begin{equation}
\label{chi} \chi=\left(\frac{\partial\delta
n}{\partial\delta\mu}\right)_{\bar{\mu},\Delta}+\left(\frac{\partial
\delta
n}{\partial\Delta}\right)^2_{\bar{\mu},\delta\mu}\left(\frac{\partial^2
\Omega}{\partial \Delta^2}\right)_{\bar{\mu},\delta\mu}^{-1}.
\end{equation}
Since $(\partial\delta n/\partial\delta\mu)_{\bar{\mu},\Delta}$ is
always positive, the stability condition $\chi>0$ is controlled by
the gap susceptibility
$\kappa_\Delta=(\partial^2\Omega/\partial\Delta^2)_{\bar{\mu},\delta\mu}$
which determines if the solution of the gap equation is the
minimum of the thermodynamic potential.

The gap susceptibility $\kappa_\Delta$ can be explicitly evaluated
as
\begin{eqnarray}
\label{kappa} \kappa_\Delta&=&2\int{d^3{\vec p}\over
(2\pi)^3}\bigg[{\Delta^2\over
(E_\Delta^-)^2}\bigg({1-f(E_1)-f(E_2)\over
E_\Delta^-}\nonumber\\
&& \ \ \ \ \ \ \ \ \ \ \ \ \ \ \ \ \ \ \ \ \ \ \ \ +f'(E_1)+f'(E_2)\bigg)\nonumber\\
&& \ \ \ \ \ \ \ \ \ \ \ \ \ +{\Delta^2\over
(E_\Delta^+)^2}\bigg({1-f(E_3)-f(E_4)\over
E_\Delta^+}\nonumber\\
&& \ \ \ \ \ \ \ \ \ \ \ \ \ \ \ \ \ \ \ \ \ \ \ \
 +f'(E_3)+f'(E_4)\bigg)\bigg],
\end{eqnarray}
where we have considered the gap equation for the condensate
$\Delta$. At weak coupling and at $T=0$, $\kappa_\Delta$ can be
evaluated as
\begin{equation}
\kappa_\Delta\approx\frac{2\bar{\mu}^2}{\pi^2}\left[1-\frac{\delta\mu\theta(\delta\mu-\Delta)}{\sqrt{\delta\mu^2-\Delta^2}}\right]
\end{equation}
which leads to $\chi<0$ in the gapless phase. Note that while the
Sarma instability $\kappa_\Delta<0$ can be cured via charge
neutrality constraint, the instability $\kappa_\Delta<0$ induced
by density fluctuations can not be removed by charge neutrality.
Similar to the Meissner mass squared, $\kappa_\Delta$ near $T_c$
takes the form $\kappa_\Delta\propto 1-T/T_c>0$. Therefore, the
uniform superfluid phase is stable against density fluctuations at
temperature close to $T_c$.

The temperature behavior of the gap susceptibility $\kappa_\Delta$
is illustrated in Fig.\ref{fig2}. In any case, there exists a
temperature window close to the critical temperature where
$\kappa_\Delta$ is positive which means that the gapless phase is
magnetically stable and hence stable against the phase separation.
Note that the stable region against the LOFF phase is larger than
the stable region against the phase separation.

\section {Neutral 2SC/g2SC Phase}
\label{s3}
We investigate in this section the two flavor color
superconductivity in an extended NJL model. The Lagrangian density
of the model including quark-quark interaction sector is defined
as
\begin{eqnarray}
\label{njl} {\cal L} &=&
\bar{\psi}\left(i\gamma^{\mu}\partial_{\mu}-m_0\right)\psi
+G_S\left[\left(\bar{\psi}\psi\right)^2+\left(\bar{\psi}i\gamma_5\vec{
\tau}\psi\right)^2 \right]\nonumber\\
&&+G_D\left(\bar\psi^C
i\gamma^5\varepsilon\epsilon^\gamma\psi\right)\left(\bar\psi
i\gamma^5\varepsilon\epsilon^\gamma\psi^C\right) \ ,
\end{eqnarray}
where $G_S$ and $G_D$ are, respectively, coupling constants in
color singlet and anti-triplet channels, the quark field
$\psi_{i\alpha}$ with flavor index $i$ and color index $\alpha$ is
a flavor doublet and a color triplet as well as a four-component
Dirac spinor, ${\bf \tau} =(\tau_1, \tau_2, \tau_3)$ are Pauli
matrices in flavor space, and
$(\varepsilon)_{ij}\equiv\varepsilon^{ij}$ and
$(\epsilon^\gamma)_{\alpha\beta}\equiv\epsilon^{\alpha\beta\gamma}$
are, respectively, totally antisymmetric tensors in flavor and
color spaces. We focus in the following on the color symmetry
breaking phase with nonzero diquark condensates defined as
\begin{eqnarray}
\label{delta} \phi_\gamma &=& -2G_D\langle
\bar\psi^Ci\gamma^5\varepsilon\epsilon^\gamma\psi\rangle,
\nonumber\\
\phi_\gamma^* &=& -2G_D\langle \bar\psi
i\gamma^5\varepsilon\epsilon^\gamma\psi^C\rangle, \ \ \ \ \
\gamma=1,2,3\ .
\end{eqnarray}
To ensure color and electric neutralities, one should introduce a
set of color chemical potentials ${\mu_a}(a=1,2,...,8)$ with
respect to color charges $Q_1,Q_2,...,Q_8$ and an electric
chemical potential $\mu_e$ with respect to the electric charge
$Q_e$. The ground state of the system is determined by minimizing
the thermodynamical potential,
$\partial\Omega/\partial\phi_\gamma=0$ under the charge neutrality
constraint $Q_e=0$ and $Q_a=0(a=1,2,...,8)$.

Since the model Lagrangian is invariant under the color $SU(3)$
transformation, we can choose a specific color symmetry breaking
direction. The most convenient choice is
\begin{equation}
\label{gauge} \phi_1=\phi_2=0\ ,\ \ \phi_3\equiv\Delta\neq 0\ .
\end{equation}
To simplify the calculation, we consider the chiral limit with
$m_0=0$ and assume chiral symmetry restoration in the color
superconducting phase. This assumption is confirmed when the
coupling constant $G_D$ in the diquark channel is not too
large\cite{mass1,mass2}. In the specific case (\ref{gauge}),
$Q_1,\cdots,Q_7$ vanish automatically and only $Q_8$ can be
nonzero. Therefore, we can introduce the color chemical potential
$\mu_8$ only, and the quark chemical potential matrix elements can
be expressed as
\begin{equation}
\mu_{ij}^{\alpha\beta}=(\mu\delta_{ij}-\mu_eQ_{ij})\delta_{\alpha\beta}+\frac{2}{\sqrt{3}}\mu_8\delta_{ij}(T_8)_{\alpha\beta}.
\end{equation}

In mean field approximation the thermodynamical potential $\Omega$
of the system can be expressed as
\begin{eqnarray}
\Omega={\Delta^2\over 4G_D}-{T\over 2}\sum_n\int{d^3 {\vec p}\over
(2\pi)^3}\textrm{Tr} \ln {\cal S}^{-1}+\Omega_e\ ,
\end{eqnarray}
where $\Omega_e$ is the contribution from the free electron gas
\begin{eqnarray}
\Omega_e=-\frac{1}{12\pi^2}\left(\mu_e^4+2\pi^2\mu_e^2T^2+\frac{7\pi^4}{15}T^4\right).
\end{eqnarray}
The inverse of the quark propagator ${\cal S}^{-1}$ in the
Nambu-Gorkov space can be written as
\begin{equation}
{\cal S}^{-1}=\left(\begin{array}{cc}
[G_0^+]^{-1}&-i\gamma_5\varepsilon\epsilon^3\Delta
\\ -i\gamma_5\varepsilon\epsilon^3\Delta
&[G_0^-]^{-1}\end{array}\right)
\end{equation}
with $[G_0^\pm]^{-1}=(i\omega_n\pm\mu_{ij}^{\alpha\beta})\gamma_0-
\vec{\gamma}\cdot\vec{p}$. After a straightforward algebra
$\Omega$ can be evaluated as
\begin{equation}
\Omega={\Delta^2\over 4G_D}-\sum_IN_I\int{d^3 {\vec p}\over
(2\pi)^3}W(E_I)+\Omega_e,
\end{equation}
where the sum runs over all quasiparticles. The quasiparticle
dispersions $E_I(p)$ calculated by $\det{\cal S}^{-1}(p)=0$ are
given by (\ref{e1234}) and
\begin{eqnarray}
&& E_5(p)= p+\mu_{u3},\ \ \ E_6(p)= p-\mu_{u3},\nonumber\\
&& E_7(p)= p+\mu_{d3},\ \ \ E_8(p)= p-\mu_{d3},
\end{eqnarray}
with the chemical potential mismatch $\delta\mu=\mu_e/2$ and
averaged chemical potential $\bar\mu = \mu-\mu_e/6+\mu_8/3$ for
paired quarks, where $\mu=\mu_B/3$ is related to the baryon
chemical potential $\mu_B$. The degenerate factor $N_I$ is $2$ for
$I=1,2,3,4$ and $1$ for $I=5,6,7,8$. Minimizing the thermodynamic
potential $\Omega$,
\begin{equation}
\label{delta}
\partial\Omega/\partial\Delta=0,
\end{equation}
and considering the charge neutrality conditions
\begin{equation}
\label{charge}
\partial\Omega/\partial\mu_8 =0,\ \ \
\partial\Omega/\partial\mu_e = 0,
\end{equation}
we can determine simultaneously the order parameter $\Delta$ and
chemical potentials $\mu_e$ and $\mu_8$ in the neutral uniform
color superconductor. At weak coupling, the color chemical
potential $\mu_8$ is only a few MeV, and the electric charge
neutrality plays the role of the condition $n_1/n_2=2$ in the toy
two-species model.
\begin{figure}[!htb]
\begin{center}
\includegraphics[width=6cm]{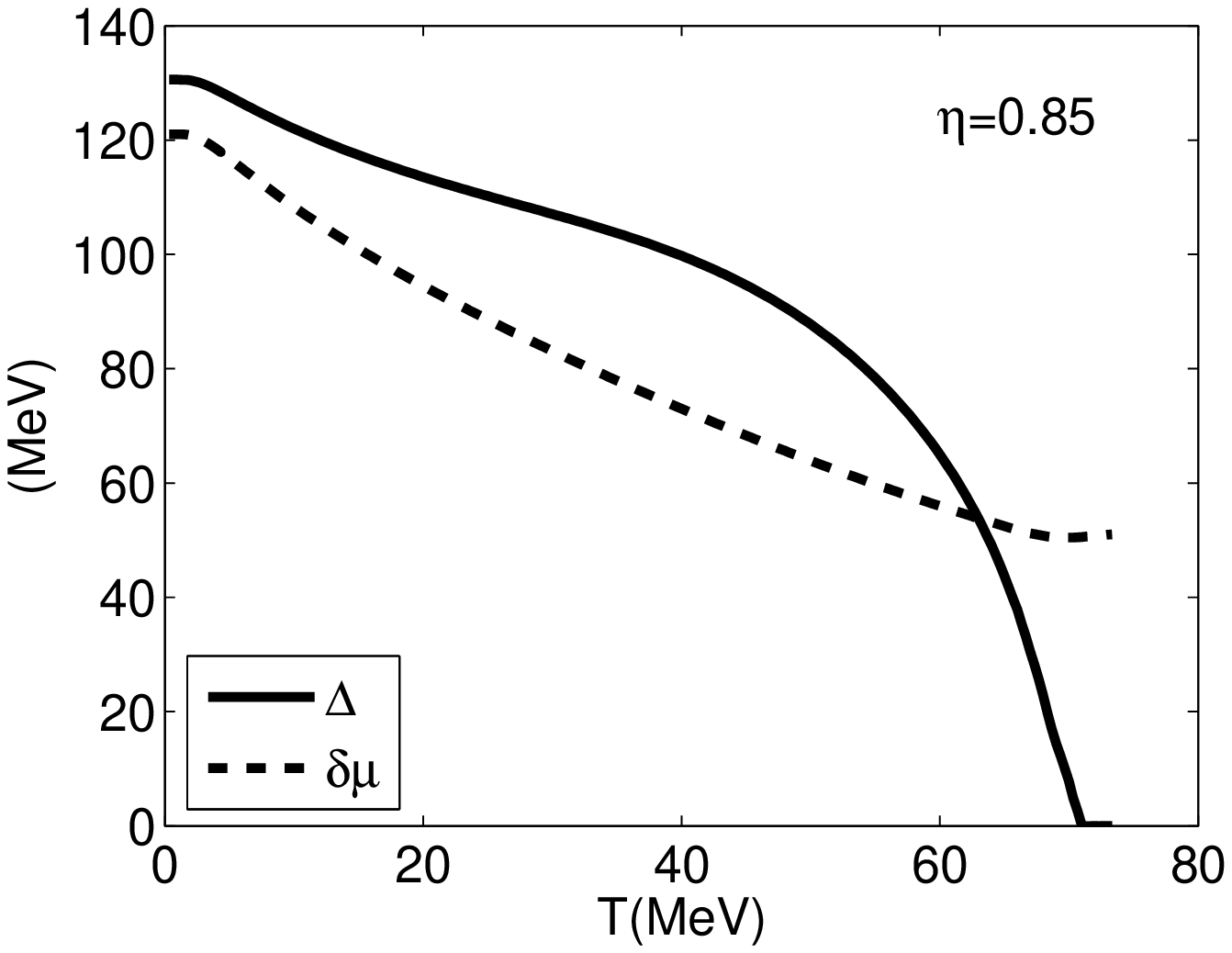}
\includegraphics[width=6cm]{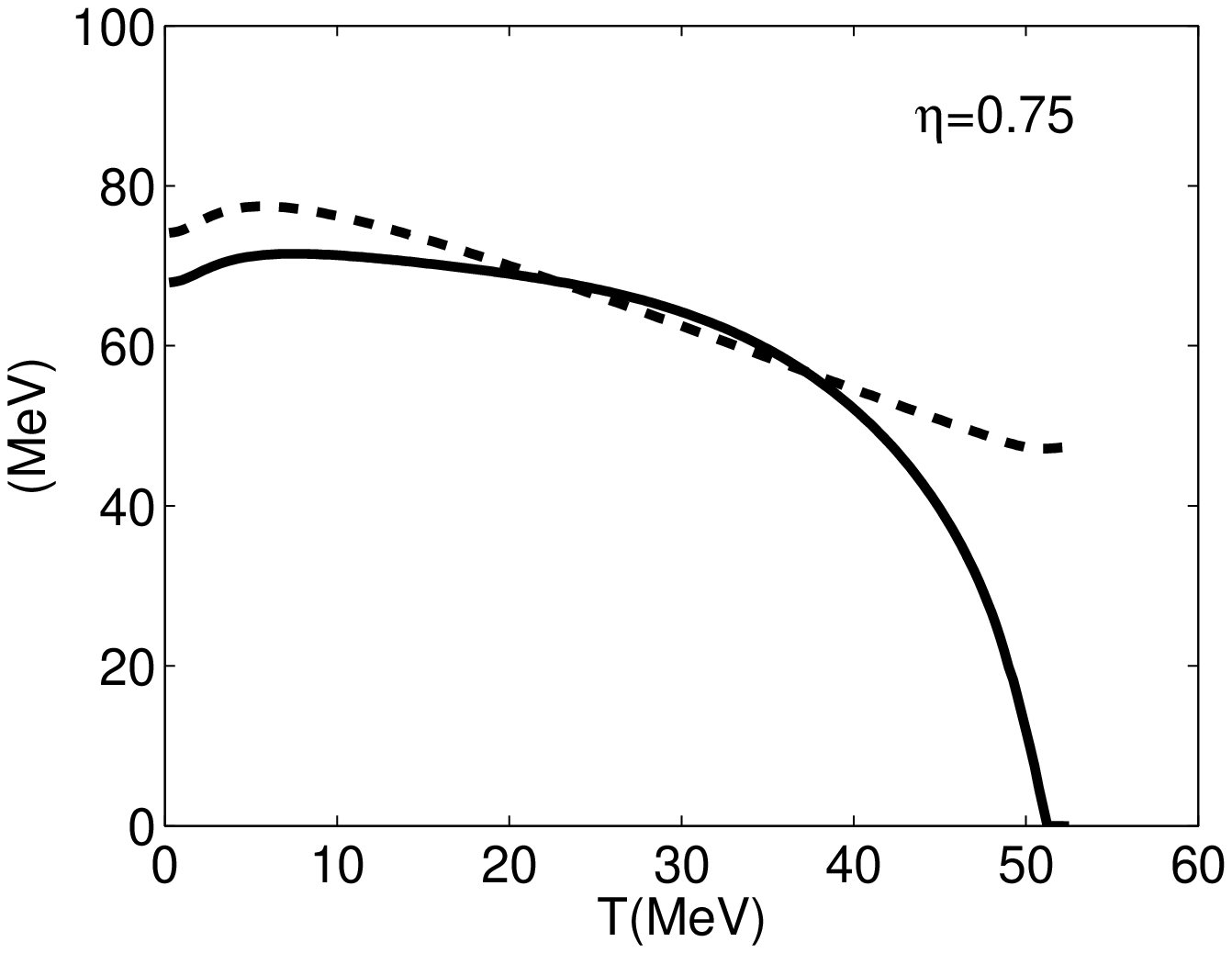}
\includegraphics[width=6cm]{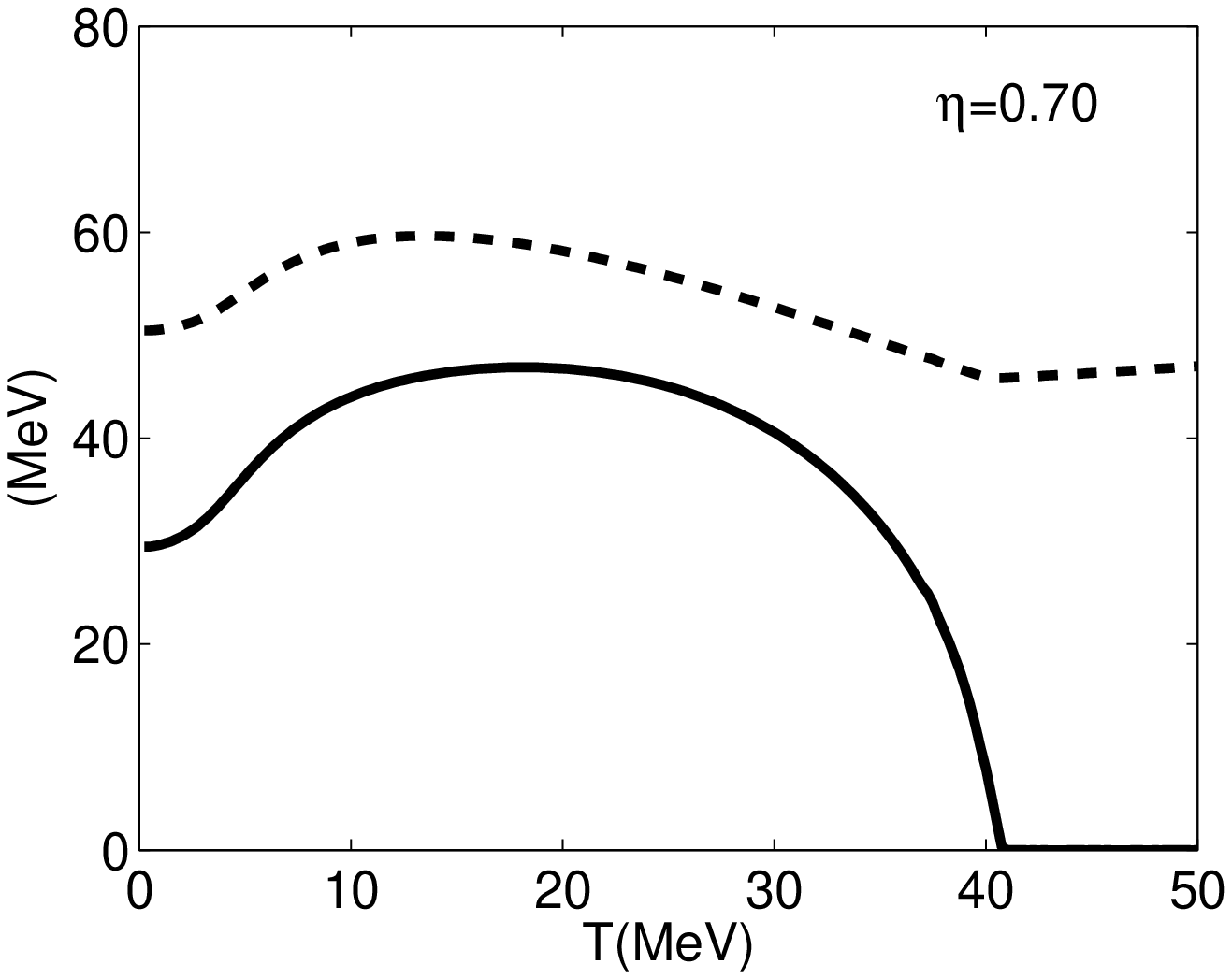}
\includegraphics[width=6cm]{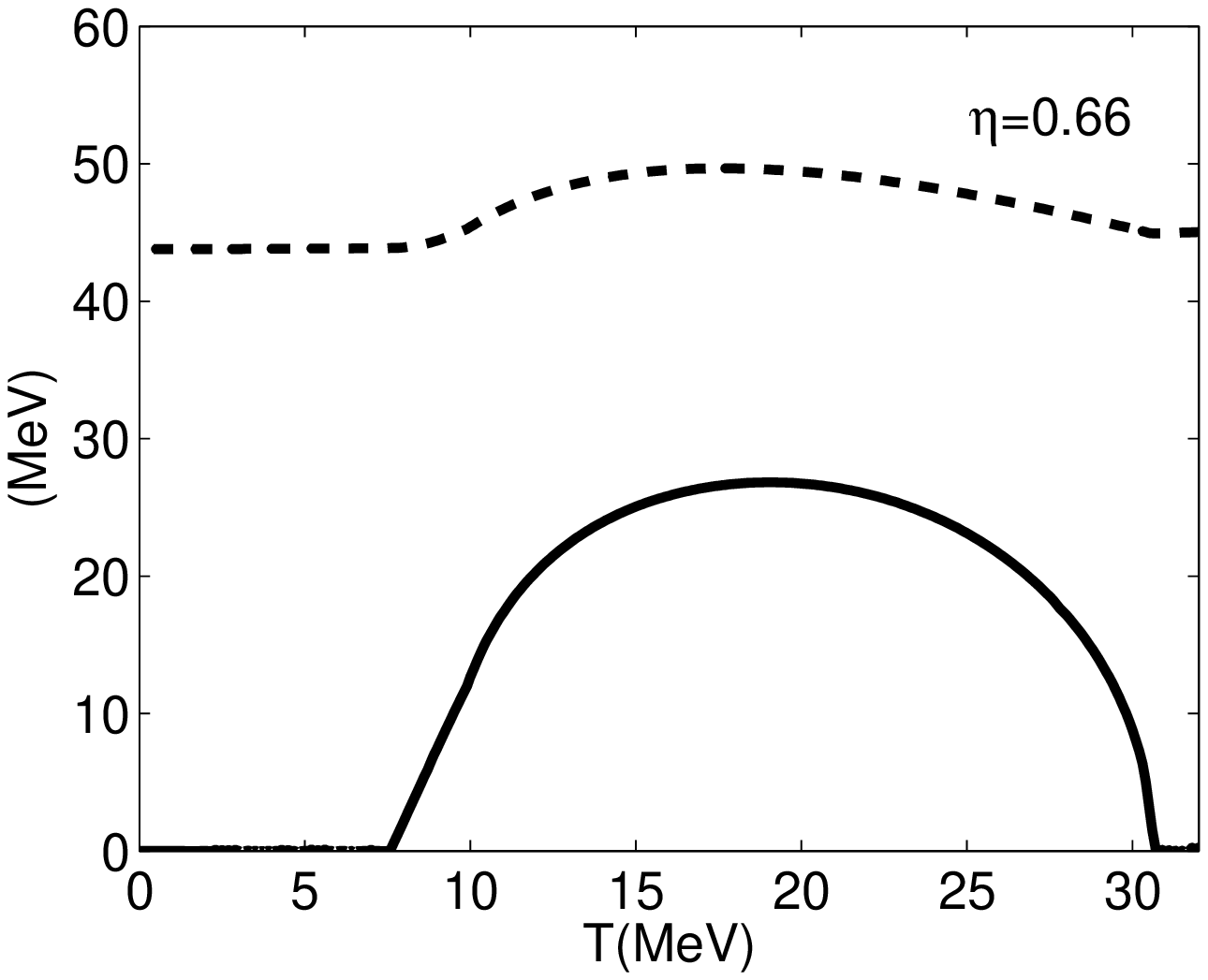}
\caption{The pairing gap $\Delta$ and chemical potential mismatch
$\delta\mu$ as functions of temperature $T$ at $\mu=400$ MeV for
four coupling values in the diquark channel. \label{fig3}}
\end{center}
\end{figure}

The numerical solutions of the condensate $\Delta$ and chemical
potential mismatch $\delta\mu$ are demonstrated in Fig.\ref{fig3}
at fixed quark chemical potential $\mu=400$ MeV for several values
of coupling $G_D$. There are three parameters in the model. The
momentum cutoff $\Lambda$ and coupling $G_S$ can be fixed by
fitting the pion decay constant and chiral condensate in the
vacuum\cite{huang}, and we denote the coupling $G_D$ by the ratio
$\eta=G_D/G_S$. For weak coupling such as $\eta=0.75, 0.70, 0.66$,
the temperature behavior of $\Delta$ and $\delta\mu$ is similar to
that in the toy two-species model. The uniform color
superconducting phase disappears at the critical coupling
$\eta_c\simeq 0.63$. For strong coupling such as $\eta=0.85$, the
quark matter is in gapped phase in a wide temperature region and
in gapless phase only in a small window close to $T_c$. Again, the
strange behavior of the gap in the low temperature region for
small $\eta$ is due to the competition of the two opposite
temperature effects for pairings with mismatched number densities.

\subsection {Meissner Mass}
The Meissner masses for gluons and photon can be evaluated via the
polarization tensor\cite{huang3}
\begin{equation}
\Pi^{\mu\nu}_{ab}(k)=\frac{T}{2}\sum_n\int{d^3{\vec p}\over
(2\pi)^3}\text {Tr}\left[\hat{\Gamma}_{a}^\mu{\cal
S}(p)\hat{\Gamma}_{b}^\nu{\cal S}(p-k)\right],
\end{equation}
where the vertex in the color-flavor space is defined as
$\hat\Gamma^\mu_a=diag\left(g_s\gamma^\mu T_a,-g_s\gamma^\mu
T_a^T\right)$ for $a=1,...,8$ and
$\hat\Gamma^\mu_a=diag\left(e\gamma^\mu Q,-e\gamma^\mu Q\right)$
for $a=\gamma$. In the 2SC/g2SC phase with the conventional choice
(\ref{gauge}), the diquark condensate breaks the gauge symmetry
group $SU(3)_c\otimes U(1)_{em}$ down to the $SU(2)_c\otimes
\tilde{U}(1)_{em}$ subgroup. Therefore, we need to calculate the
Meissner masses for the 4-8th gluons and the photon only. Most of
the analytic calculation is presented in \cite{huang3}. With the
function $A, A', B, C, D, H, J$ listed in Appendix {\ref{app}, we
obtain
\begin{eqnarray}
\label{meissner1}
m^2_{88}&=&\frac{4\alpha_s}{9\pi}\int
dpp^2\left[A+A^\prime-2B+4(C-D)\right],\\
m^2_{\gamma\gamma}&=&\frac{4\alpha_e}{9\pi}\int
dpp^2\left[4A+A^\prime+4B+2(5C+4D)\right],\nonumber\\
m^2_{8\gamma}&=&\frac{8\sqrt{\alpha_e\alpha_s}}{9\sqrt{3}\pi}\int
dpp^2\left[2A-A^\prime-B+2(C-D)\right]\nonumber
\end{eqnarray}
with $\alpha_s=g_s^2/(4\pi)$ and $\alpha_e=e^2/(4\pi)$ for the 8th
gluon and photon and
\begin{equation}
\label{meissner2}
m_4^2\equiv
m^2_{44}=m_{55}^2=m_{66}^2=m_{77}^2=\frac{4\alpha_s}{3\pi}\int
dpp^2(H+2J)
\end{equation}
for the 4-7th gluons. Due to the nonzero $m_{8\gamma}^2$, the 8th
gluon and the photon mix with each other, and the physical
Meissner masses squared are given by the eigen values
\begin{eqnarray}
m^2_{\tilde{8}\tilde{8}}=\frac{1}{2}\left[m_{88}^2+m_{\gamma\gamma}^2+\sqrt{(m_{88}^2-m_{\gamma\gamma}^2)^2+4m_{8\gamma}^4}\right],&&\nonumber\\
m^2_{\tilde{\gamma}\tilde{\gamma}}=\frac{1}{2}\left[m_{88}^2+m_{\gamma\gamma}^2-\sqrt{(m_{88}^2-m_{\gamma\gamma}^2)^2+4m_{8\gamma}^4}\right].&&
\end{eqnarray}
When the coupling is not very large, we have approximately
$m_{\tilde{\gamma}\tilde{\gamma}}^2\simeq0$, which is consistent
with the analysis of symmetry breaking in weak coupling limit.
Since $\alpha_s\sim 1 >> \alpha_e\simeq 1/137$, we have
\begin{equation}
m_8^2\equiv m^2_{\tilde{8}\tilde{8}}\simeq m^2_{88}.
\end{equation}
With the explicit form of the functions $A,A^\prime,B,C,D$ shown
in Appendix \ref{app}, $m_8^2$ takes the same expression as
$m_A^2$ in the toy two-species model,
\begin{equation}
m^2_8=\frac{2}{3}m_A^2,
\end{equation}
and therefore, $m_8^2$ and $m_A^2$ have exactly the same
temperature behavior. Now the most important task is to
investigate whether $m_4^2$ can be positive at finite temperature.
For $T\rightarrow T_c$, we have $\mu_8\rightarrow 0$ and the
functions $H$ and $J$ are reduced to
\begin{equation}
H=\sum_{i=u}^d\left[f(p-\mu_i)+f(p+\mu_i)\right],\ \ \ J=H/p
\end{equation}
with $\mu_u=\mu-2\mu_e/3$ and $\mu_d=\mu+\mu_e/3$. Employing the
same trick used in Section \ref{s2}, we find $m_4^2=0$ for $T\geq
T_c$ and
\begin{equation}
m_4^2(T) \propto1-T/T_c,\ \ \ T\rightarrow T_c
\end{equation}
for $T$ below but close to $T_c$. Therefore, there must exist a
temperature window near the critical temperature where the two
flavor color superconductor is free from chromomagnetic
instability.
\begin{figure}[!htb]
\begin{center}
\includegraphics[width=6cm]{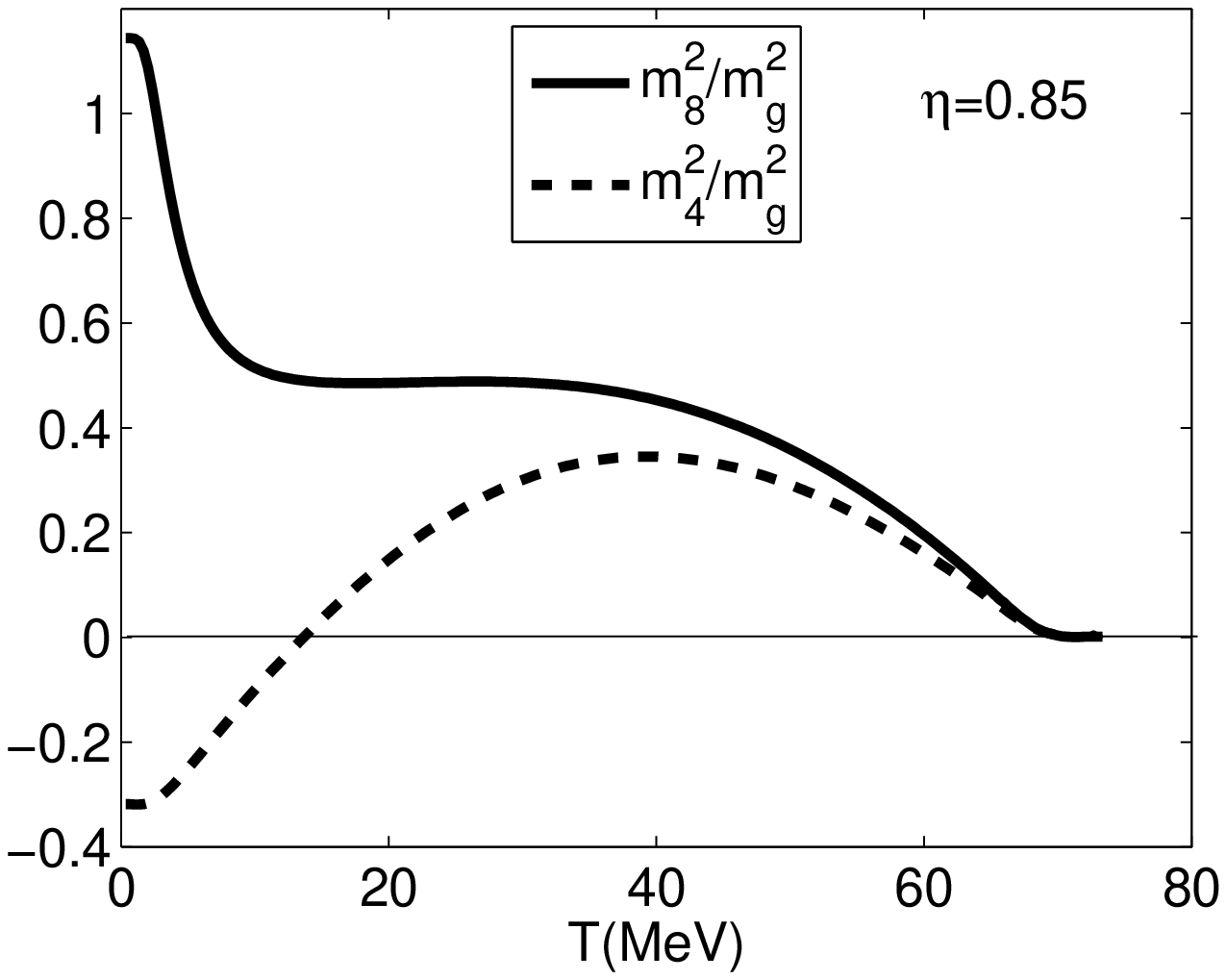}
\includegraphics[width=6cm]{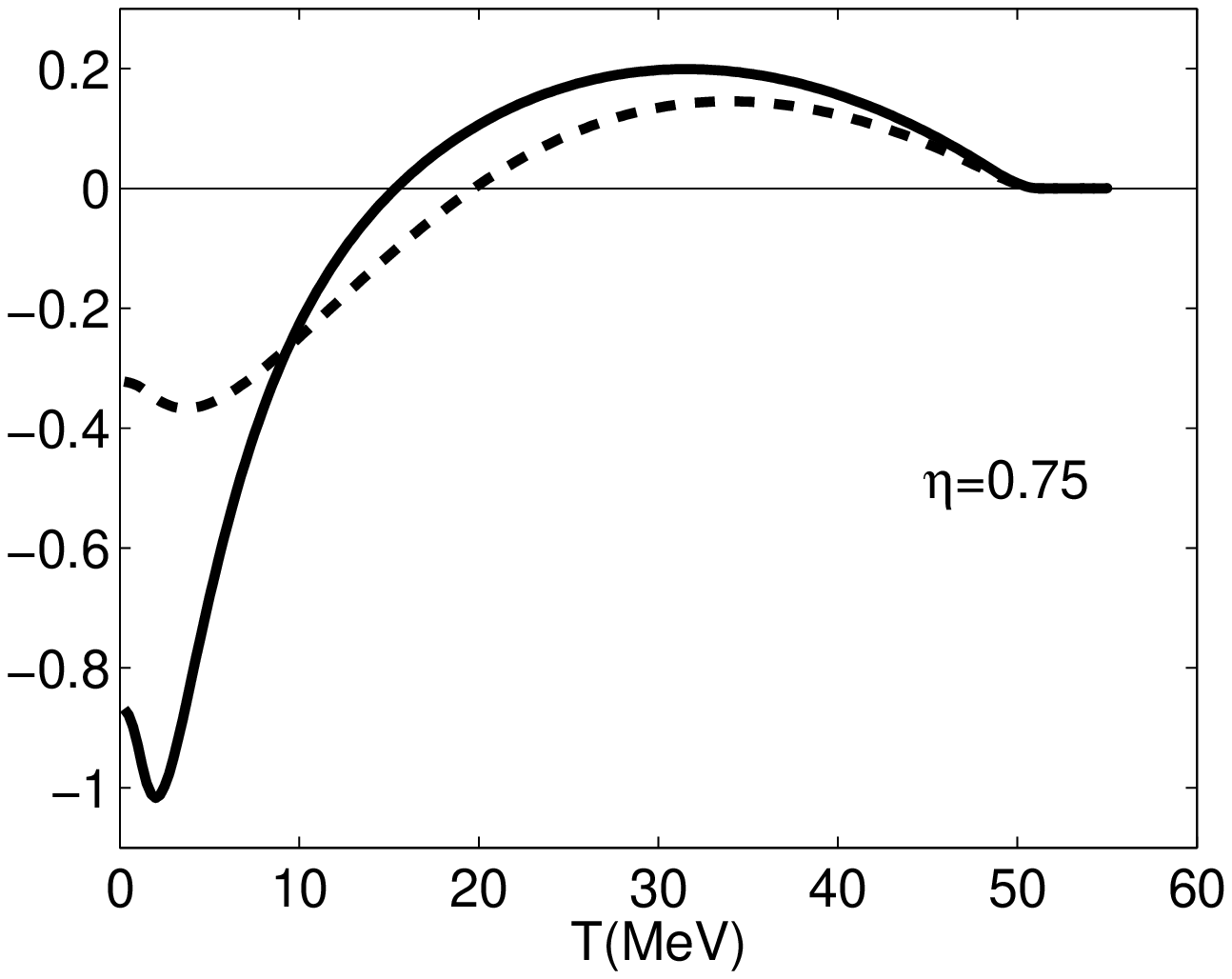}
\includegraphics[width=6cm]{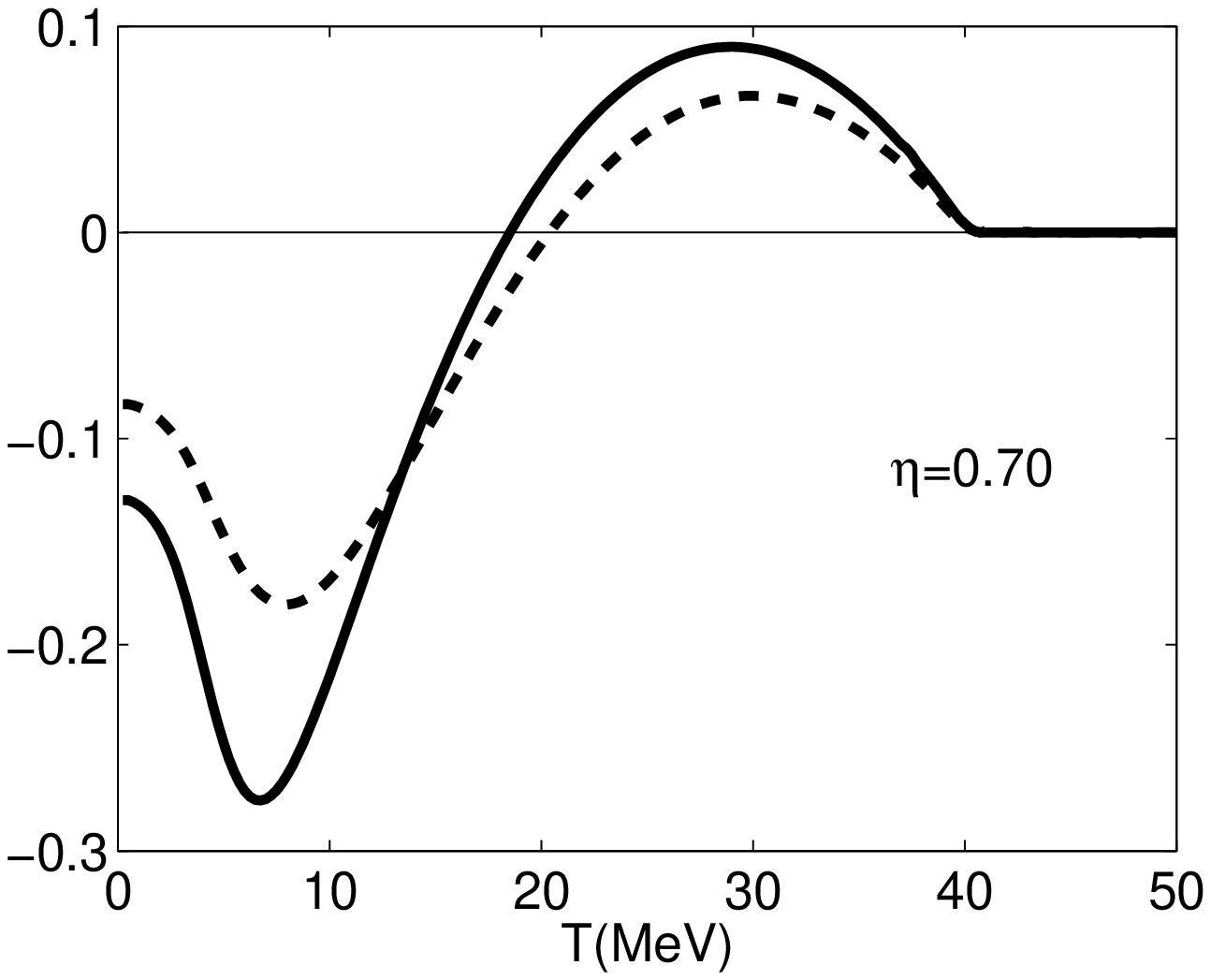}
\includegraphics[width=6cm]{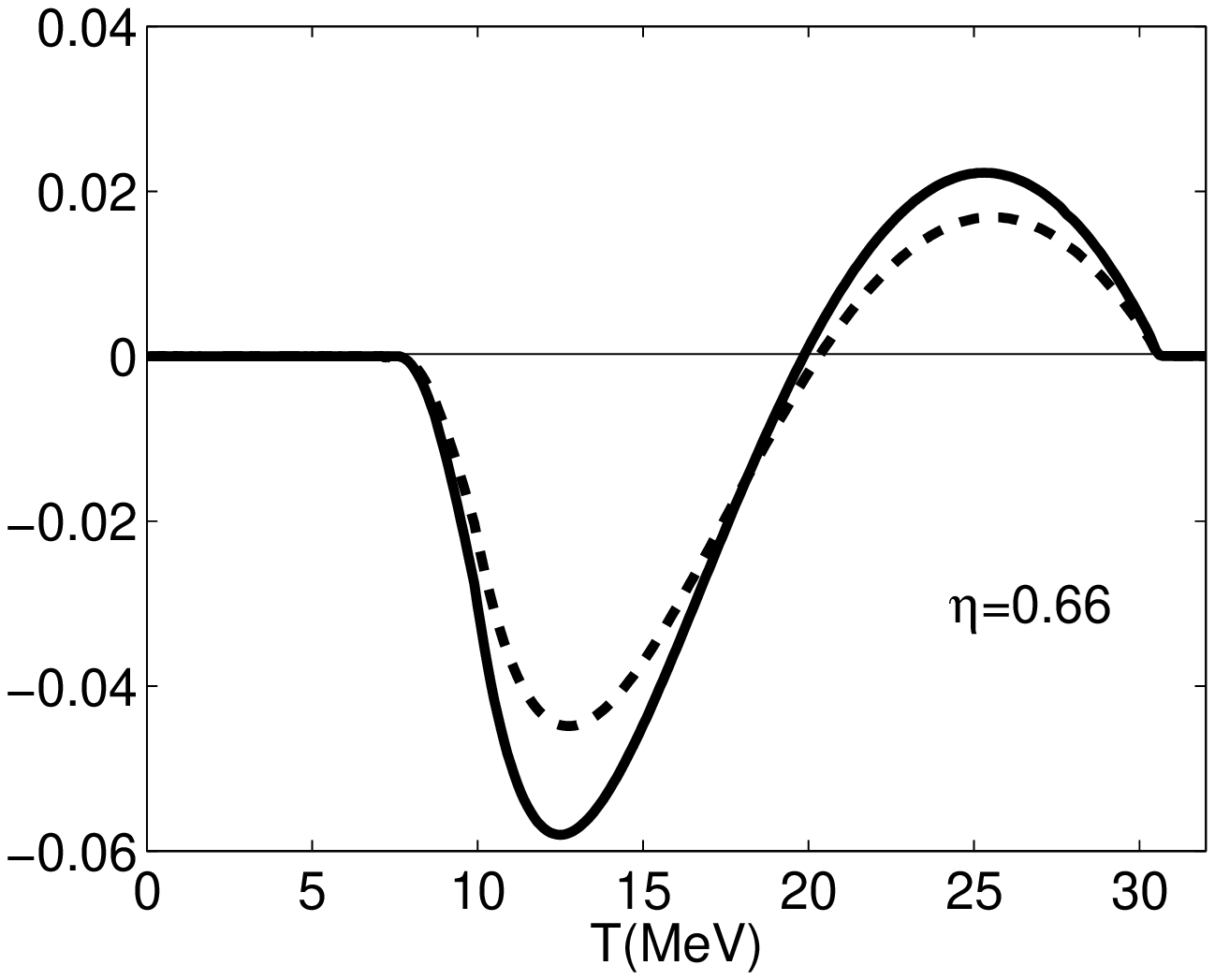}
\caption{The Meissner masses squared $m_4^2$ and $m_8^2$ scaled by
$m_g^2=4\alpha_s\bar{\mu}^2/3\pi$ as functions of temperature $T$
at $\mu=400$ MeV for four coupling values. \label{fig4}}
\end{center}
\end{figure}

In Fig.\ref{fig4} we illustrate the temperature behavior of the
Meissner masses squared $m_4^2$ and $m_8^2$ at fixed chemical
potential $\mu=400$ MeV for several coupling values. For weak
coupling $0.62<\eta<0.8$, there exist two intermediate
temperatures $T_4$ and $T_8$, $m_4^2$ is negative at $0<T<T_4$ and
positive at $T_4<T<T_c$, and $m_8^2$ is negative at $0<T<T_8$ and
positive at $T_8<T<T_c$. In a wide range of coupling, $T_4$ is
larger than $T_8$. Only for sufficiently small coupling
$\eta<0.66$, $T_4$ coincides with $T_8$ or even becomes less than
$T_8$. For strong coupling $\eta>0.8$, the 8th gluon is free from
chromomagnetic instability at any temperature, but the 4-7th
gluons suffer negative Meissner mass squared in the low
temperature region $0<T<T_4$ which indicates the instability
against the gluonic phase\cite{gorbar}.

\subsection {Density Fluctuation}
Like the toy two-species model, we can study the stability of
color superconductivity against the number fluctuations of the
paired quarks\cite{Iida}. Similarly, the stability condition can
be reduced to $\chi=\left(\partial\delta
n/\partial\delta\mu\right)_{\mu}>0$ where $\delta
n=n_{d1}-n_{u2}=n_{d2}-n_{u1}$ is the density imbalance between
the paired quarks.  Using the techniques in Section II, the gap
susceptibility $\kappa_\Delta$ which controls the stability takes
almost the same form (\ref{kappa}), the only difference is the
replacement of the factor of 2 in front of the momentum
integration in (\ref{kappa}) by the factor of 4.

In Fig.\ref{fig5} we show the temperature behavior of the gap
susceptibility $\kappa_\Delta$ for four coupling values in the
diquark channel. In any case, there indeed exists a temperature
window where $\kappa_\Delta$ is positive and therefore the gapless
phase is magnetically stable and hence stable against the phase
separation. Recently, it is argued that the Higgs
instability\cite{higgs} indicates the instability of 2SC/g2SC
against the mixed phase. The gap susceptibility is the long
wavelength limit of the Higgs instability. For a complete study we
may need to check the Higgs instability at finite temperature.
\begin{figure}[!htb]
\begin{center}
\includegraphics[width=6cm]{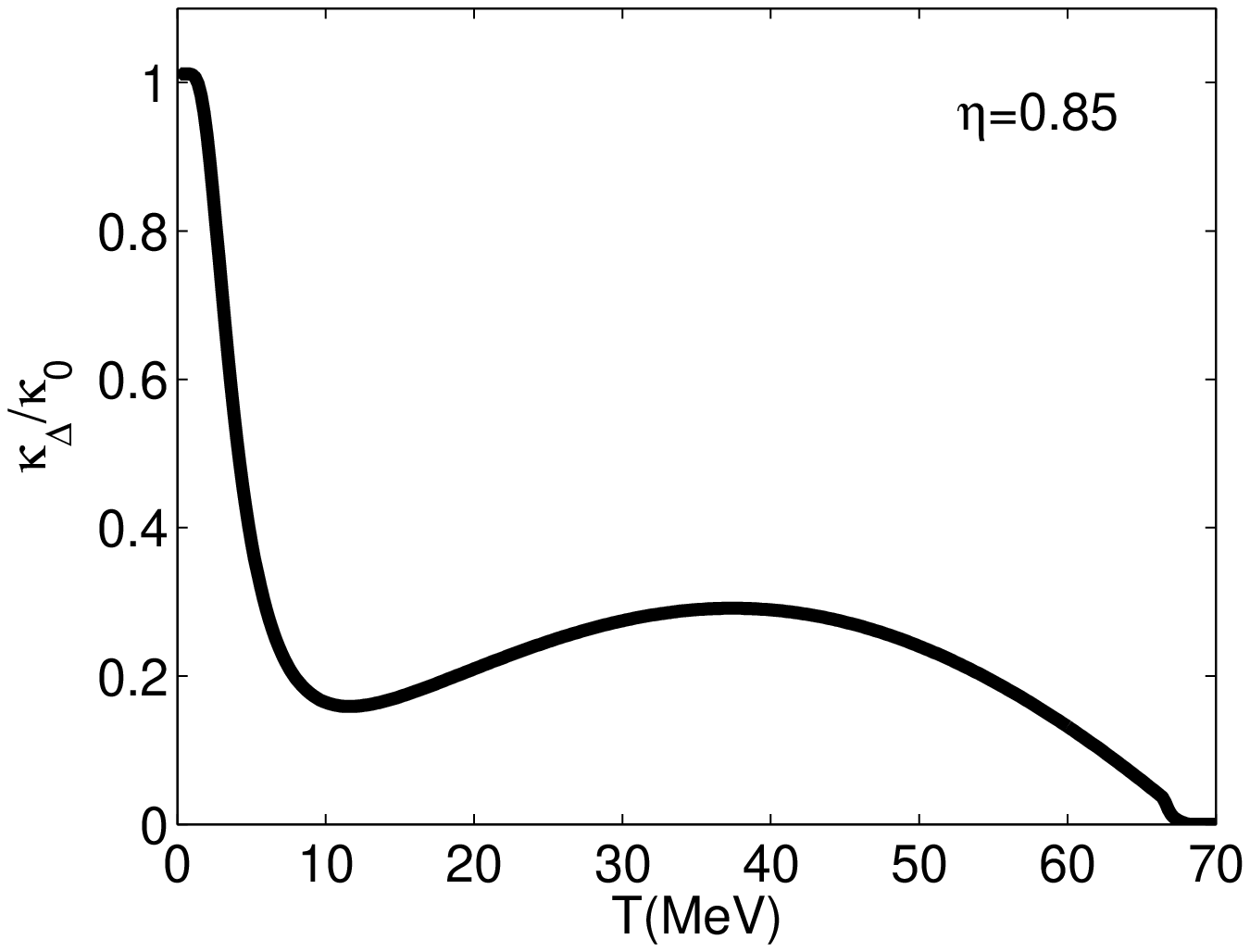}
\includegraphics[width=6cm]{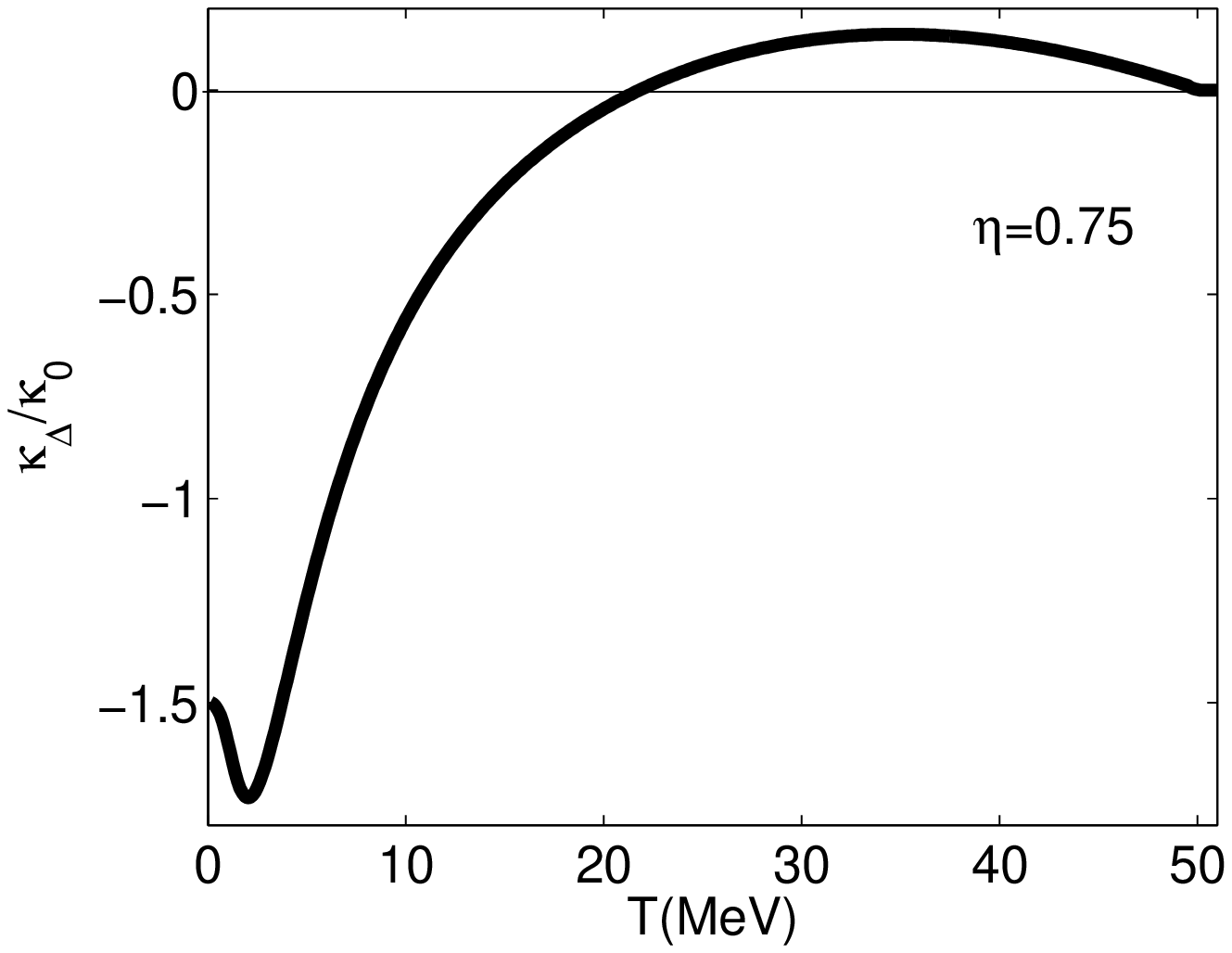}
\includegraphics[width=6cm]{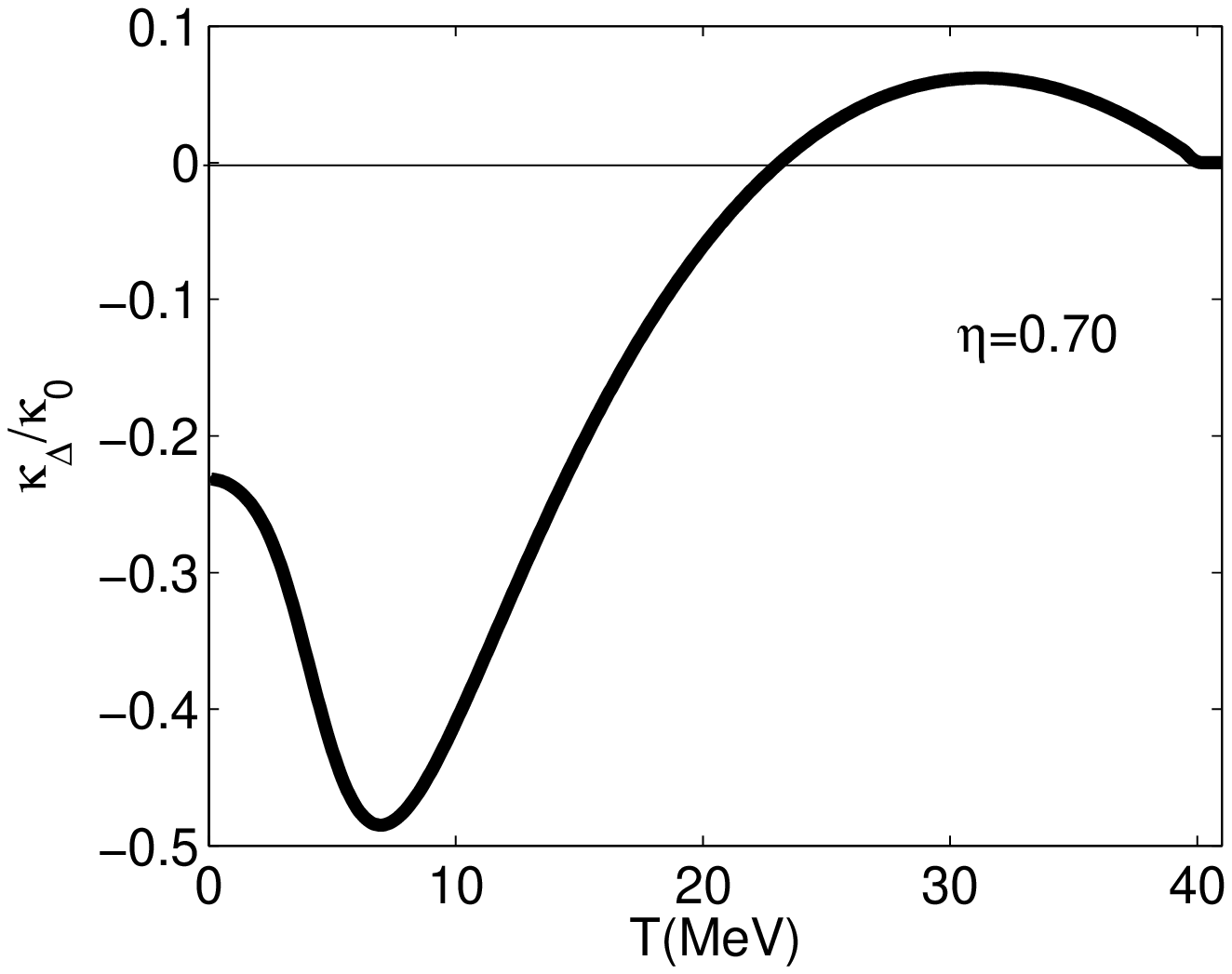}
\includegraphics[width=6cm]{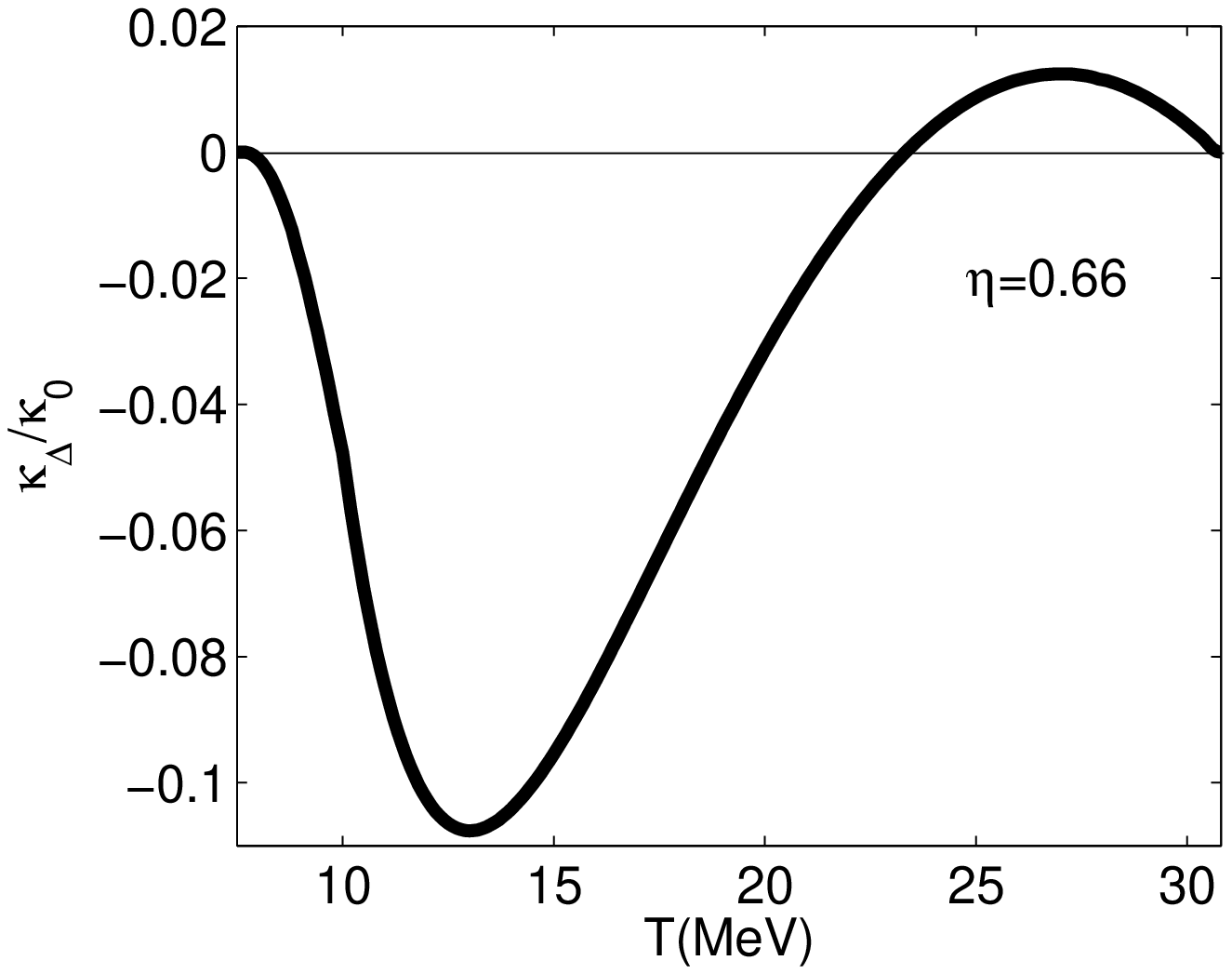}
\caption{The gap susceptibility $\kappa_\Delta$ scaled by
$\kappa_0=2\bar{\mu}^2/\pi^2$ as a function of temperature $T$ at
$\mu=400$MeV for four different couplings. \label{fig5}}
\end{center}
\end{figure}

\section {Inhomogeneous Phases}
\label{s4}
We have calculated the Meissner masses squared  and the gap
susceptibility in the neutral 2SC/g2SC phase, and found that there
exists a temperature window close to the critical temperature
$T_c$ where they are both positive. Therefore, the uniform phase
is free from the chromomagnetic instability and stable against the
mixed phase in this temperature region. In general case, the
stable region for the 4-7th gluons is smaller than the stable
region for the 8th gluon, namely $T_4>T_8$. Only for sufficiently
small coupling $\eta<0.66$, the 4-7th gluons can have larger
stable region than the 8-th gluon, namely $T_4\le T_8$. For large
enough coupling, the 8th gluon is free from the chromomagnetic
instability at any temperature and there exists only a turning
temperature $T_4$ where $m_4^2$ changes sign.

If $T_4>T_8$, the single-plane wave LOFF phase can not completely
cure the chromomagnetic instability, since there exists a region
$T_4<T<T_8$ where only $m_4^2$ is negative. In this section we
will focus on possible inhomogeneous phases in two limits, very
weak coupling and very strong coupling. For very weak coupling, we
have $T_4<T_8$, it has been shown in \cite{gorbar2} that the
neutral LOFF phase is free from the chromomagnetic instability at
least at $T=0$. For very strong coupling, only 4-7th gluons
suffers instability, the introduction of gluonic phase may be
sufficient to cure the instability.

\subsection {LOFF Phase}
We firstly discuss the LOFF state. The numerical calculation will
be performed at very weak coupling $\eta\leq 0.66$. The order
parameters $\phi_\gamma$ and $\phi_\gamma^*$ are complex conjugate
to each other and can be set to be real only in a uniform color
superconductor. In the LOFF phase the phase factor of the order
parameter is non-uniform. For the sake of simplicity we take the
following single wave LOFF ansatz
\begin{equation}
\phi_3= \Delta e^{2i\vec{q}\cdot\vec{x}},\ \ \ \phi_3^* = \Delta
e^{-2i\vec{q}\cdot\vec{x}}.
\end{equation}
It has been shown that in the two flavor color superconductor the
LOFF momentum $\vec{q}$ can be regarded as the 8th gluon
condensate\cite{gorbar2} or a spontaneously generated
Nambu-Goldstone current\cite{huang4}, namely
\begin{equation}
\vec{q}={1\over 2\sqrt
3}g\langle\vec{A}_8\rangle=\langle\vec{\nabla}\varphi_8\rangle.
\end{equation}
With a transformation for the quark fields,
\begin{equation}
\chi(\vec{x})=\psi(\vec{x})e^{-i\vec{q}\cdot\vec{x}},\ \ \
\chi^C(\vec{x})=\psi^C(\vec{x})e^{i\vec{q}\cdot\vec{x}},
\end{equation}
we can express the thermodynamic potential as
\begin{equation}
\Omega ={\Delta^2\over 4G_D}-{T\over 2}\sum_n\int{d^3 \vec{p}\over
(2\pi)^3}\text{Tr} \ln {\cal S}_q^{-1}(i\omega_n,\vec{p})
\end{equation}
with the inverse of the mean field propagator in LOFF state
\begin{equation}
{\cal S}_q^{-1}(i\omega_n,\vec{p})=\left(\begin{array}{cc} [
G_q^+]^{-1}&-i\gamma_5\varepsilon\epsilon^3\Delta
\\ -i\gamma_5\varepsilon\epsilon^3\Delta&[G_q^-]^{-1}\end{array}\right),
\end{equation}
where the $q$ dependent free propagators are defined as
\begin{equation}
[G_q^\pm]^{-1}=(i\omega_n\pm\hat{\mu})\gamma_0-
\vec{\gamma}\cdot(\vec{p}\pm\vec{q})\ .
\end{equation}
The neutral LOFF state should be determined self-consistently by
the gap equations for the condensate and pair momentum and the
charge neutrality constraints,
\begin{equation}
\label{4omega}
\frac{\partial\Omega}{\partial\Delta}=0,\ \
\frac{\partial\Omega}{\partial q}=0,\ \
\frac{\partial\Omega}{\partial\mu_e}=0,\ \
\frac{\partial\Omega}{\partial\mu_8}=0,
\end{equation}
where we have chosen a suitable frame with $\vec{q}=(0,0,q)$.
Since $\mu_8$ is very small in weak coupling we will simply set
$\mu_8=0$ in numerical calculations.

Following the treatment in \cite{giannakis}, we expand the
thermodynamic potential in powers of the pair momentum $q$ in the
vicinity of $q=0$,
\begin{equation}
\Omega(q)-\Omega(0)=\frac{\partial \Omega}{\partial
q}\bigg|_{q=0}q+\frac{1}{2}\frac{\partial^2 \Omega}{\partial
q^2}\bigg|_{q=0}q^2+\cdots.
\end{equation}
The linear term vanishes automatically because of the gap equation
for $q$. Using the relation
\begin{equation}
{\cal S}^{-1}(\vec{q})={\cal
S}^{-1}(\vec{q}=0)-\tau_3\vec{\gamma}\cdot\vec{q},
\end{equation}
and taking the derivative expansion of $\Omega$ in powers of the
gauge field, we can easily obtain the relation between the
momentum susceptibility $\kappa_q$ and the Meissner mass squared
for the 8th gluon
\begin{equation}
\label{kq}
\kappa_q=\partial^2 \Omega/\partial
q^2\big|_{q=0}=12m_{8}^2/g^2.
\end{equation}
On the other hand, since $q=0$ must be a solution of the gap
equation $\partial \Omega/\partial q=0$ which corresponds to the
homogeneous 2SC/g2SC phase, the first order derivative of $\Omega$
with respect to $q$ must take the form\cite{he6}
\begin{equation}
\partial \Omega/\partial q=qQ(q).
\end{equation}
The momentum solution for the LOFF state is given by $Q(q)=0$. The
case here is similar to the gap equation for the pairing gap
$\Delta$ which contains a trivial solution $\Delta=0$
corresponding to the normal phase and a finite solution $\Delta\ne
0$ corresponding to the superfluid phase. For the formal proof
here we do not need the explicit function $Q(q)$.  From the
identity
\begin{equation}
\partial^2\Omega/\partial q^2=Q(q)+qQ^\prime(q)
\end{equation}
and (\ref{kq}), we obtain
\begin{equation}
m_{8}^2=g^2Q(0)/12,
\end{equation}
which means $Q(0)=0$ at $T=T_8$ where $m_8^2$ changes sign.
Therefore, the LOFF momentum $q$ must vanish at $T=T_8$, providing
that the neutral LOFF solution is unique for a given $T$ and
$\mu$. This indicates that there is no neutral LOFF solution at
$T>T_8$. Below but near the temperature $T_8$, the LOFF momentum
$q$ is very small, the small $q$ expansion is valid
\begin{equation}
\Omega(q)-\Omega(0)=6m_{8}^2q^2/g^2.
\end{equation}
For negative $m_{8}^2$ below $T_8$, the neutral LOFF state has
lower free energy than the uniform state at least at $T\lesssim
T_8$. Since the small $q$ expansion is like a Ginzburg-Landau
expansion, we conclude that the LOFF momentum near the critical
point behaviors as
\begin{equation}
q(T) \sim \left(1-T/T_8\right)^{1/2},\ \ \ T\rightarrow T_8.
\end{equation}

Now we turn to the numerical calculation of neutral LOFF state. The
explicit form of $\Omega$ can be obtained if we neglect the mixing
between the particles and anti-particles\cite{gorbar2} since we have
$\Delta,q\ll\bar{\mu}$ at weak coupling. It reads
\begin{equation}
\Omega={\Delta^2\over 4G_D}-\int{d^3 {\vec p}\over
(2\pi)^3}\left[\sum_{I=1}^{8}W(E_I)+\frac{1}{2}\sum_{I=9}^{16}W(E_I)\right]+\Omega_e,
\end{equation}
where the sum runs over all quasiparticles. The quasiparticle
dispersions $E_I(\vec{p})$ calculated by $\det{\cal S}_q^{-1}=0$ are
given by
\begin{eqnarray}
\label{e1to8} E_{1,2}(\vec{p}) &=& E_{\Delta q}^- +\delta
E_q\pm\delta\mu,\nonumber\\
E_{3,4}(\vec{p}) &=& E_{\Delta q}^- -\delta
E_q\pm\delta\mu,\nonumber\\
E_{5,6}(\vec{p}) &=& E_{\Delta q}^+ +\delta E_q\pm\delta\mu,\nonumber\\
E_{7,8}(\vec{p}) &=& E_{\Delta q}^+ -\delta E_q\pm\delta\mu,\nonumber\\
E_{9,10}(\vec{p}) &=& |\vec{p}+\vec{q}|-\bar{\mu}+\mu_8\pm\delta\mu,\nonumber\\
E_{11,12}(\vec{p}) &=& |\vec{p}-\vec{q}|-\bar{\mu}+\mu_8\pm\delta\mu,\nonumber\\
E_{13,14}(\vec{p}) &=& |\vec{p}+\vec{q}|+\bar{\mu}-\mu_8\pm\delta\mu,\nonumber\\
E_{15,16}(\vec{p}) &=&
|\vec{p}-\vec{q}|+\bar{\mu}-\mu_8\pm\delta\mu.
\end{eqnarray}
where $E_{\Delta q}^\pm$ and $\delta E_q$ are defined as
\begin{eqnarray}
E_{\Delta
q}^\pm&=&\sqrt{\left(|\vec{p}+\vec{q}|+|\vec{p}-\vec{q}|-2\bar{\mu}\right)^2/4+\Delta^2},\nonumber\\
\delta
E_q&=&\frac{1}{2}\left(|\vec{p}+\vec{q}|+|\vec{p}-\vec{q}|\right).
\end{eqnarray}
Notice that in our ultra-relativistic system, the thermodynamic
potential of the LOFF state suffers from a unphysical term $\sim
q^2\Lambda^2$\cite{gorbar2}. In fact, $\Omega(\Delta,q)$ should
recover the result for the free quark gas when $\Delta\rightarrow
0$. Thus to study the neutral LOFF state, we define the following
subtracted thermodynamic potential
\begin{eqnarray}
\Omega_{\text{sub}}(\Delta,q)=\Omega(\Delta,q)-\Omega(0,q)+\Omega(0,0).
\end{eqnarray}
Thus we can set the momentum $\vec{q}$ to be zero in the dispersion
of quasiparticles $9-16$ and treat the UV divergence for the
quasiparticles $1-8$. It is obvious that we recover the
thermodynamic potential of the 2SC phase when $q=0$ and the
thermodynamic potential of the free quark gas when $\Delta=0$. We
can further make approximation on $E_{\Delta q}^\pm$ and $\delta
E_q$ since $\Delta,q\ll\bar{\mu}$ at weak coupling and all integrals
are dominated over the Fermi surface $|\vec{p}|=\bar{\mu}$. In this
approximation we have
\begin{eqnarray}
E_{\Delta q}^\pm&\simeq&E_\Delta^\pm+\frac{q^2}{2p}\frac{E_0^\pm}{E_\Delta^\pm}(1-\cos^2\theta),\nonumber\\
\delta E_q&\simeq&q\cos\theta,
\end{eqnarray}
with $\theta$ being the angle between $\vec p$ and $\vec q$. In this
case the subtraction term at $T=0$ can be analytically evaluated as
\begin{equation}
\Omega_\Lambda=8\int{d^3 {\vec p}\over
(2\pi)^3}(1-\cos^2\theta)\frac{q^2}{2p}=\frac{2\Lambda^2q^2}{3\pi^2}
\end{equation}
which is from the kinetic term of the quasiparticle dispersion and
is indeed diverges as $\sim\Lambda^2q^2$. At finite temperature,
however, we do not find an analytical expression for
$\Omega_\Lambda$.

In the numerical calculation of the neutral LOFF state at finite
temperature, we then use the subtracted thermodynamic potential
$\Omega_{\text{sub}}(\Delta,q)$. The gap parameter $\Delta$ and pair
momentum $q$ as well as the electron chemical potential $\mu_e$ in
the neutral LOFF state are determined by self-consistently solving
the coupled set of equations (\ref{4omega}) but with the subtracted
thermodynamic potential $\Omega_{\text{sub}}$. For simplicity, we
will neglect the color neutrality condition and set $\mu_8=0$.

In Fig.\ref{fig6} we display the numerical result of neutral LOFF
state for $\eta=0.66$ and $0.70$. The neutral LOFF solution exists
only at low temperature, namely at $T<T_8$, and the LOFF momentum
approaches to zero continuously at $T=T_8$.  Since the LOFF phase
has lower thermodynamic potential than the uniform phase (we
checked this numerically), the LOFF phase is energetically more
favored than the uniform superconductivity. Considering both the
stable LOFF state at low temperature and the stable uniform
superconductivity at high temperature, the strange intermediate
temperature superconductivity disappears, and the order parameter
becomes a monotonously decreasing function of temperature, like in
the conventional BCS case. Especially, for $\eta=0.66$, the
uniform superconductivity does not appear at $T=0$, but LOFF phase
starts at $T=0$. For sufficiently weak coupling, the uniform
superconductivity disappears at any temperature, but the LOFF
phase can survive at low temperature. In Fig.\ref{fig7} we show
the neutral LOFF state for $\eta=0.63$ where the uniform
superconductivity starts to disappear. The temperature behavior of
the order parameter is similar to that in the conventional BCS
case.

We conclude that when LOFF phase is taken into account, the phase
diagram\cite{phase1,phase2,phase3,phase4,phase5} of neutral quark
matter is significantly changed.
\begin{figure}[!htb]
\begin{center}
\includegraphics[width=6cm]{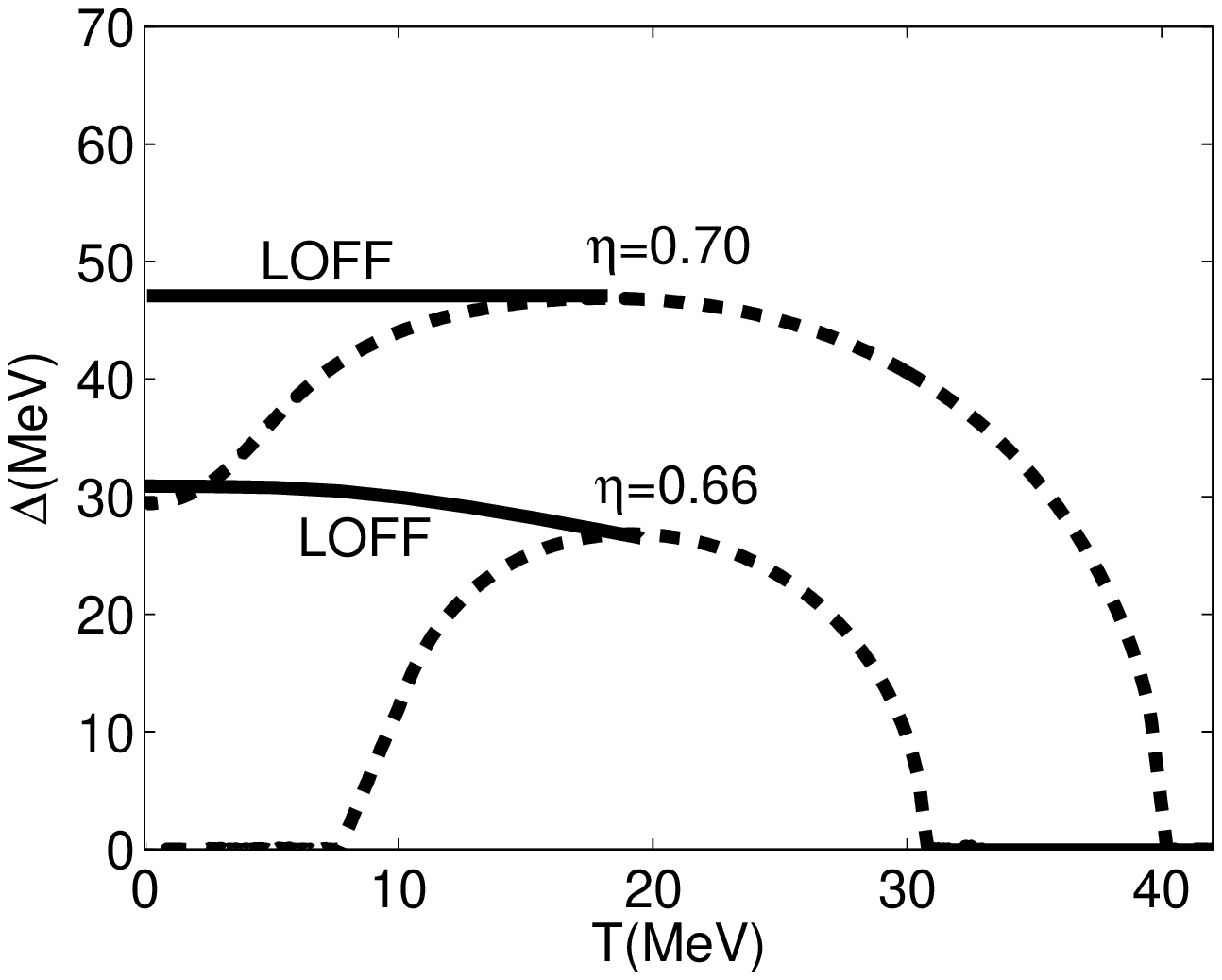}
\includegraphics[width=6cm]{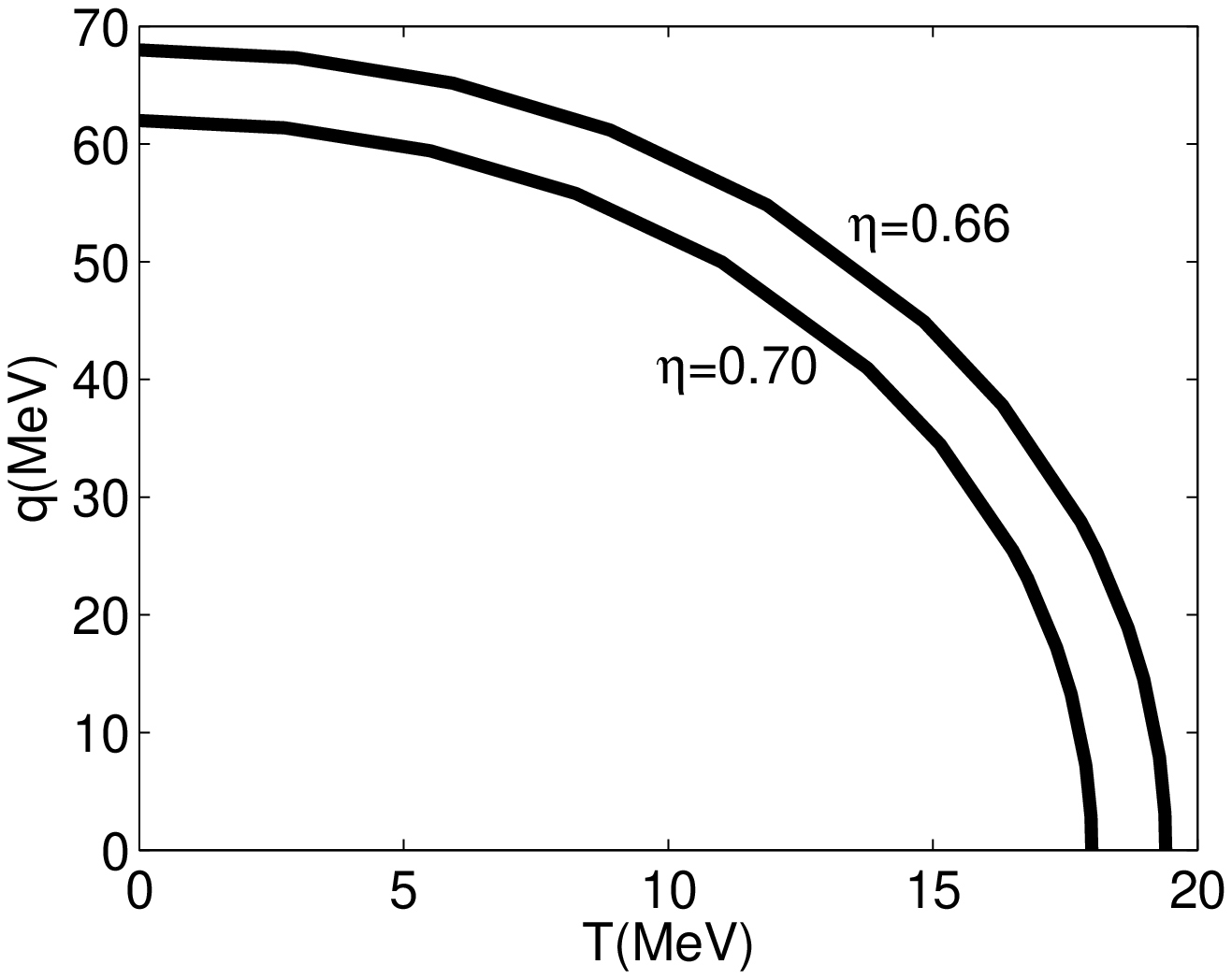}
\caption{The pairing gap $\Delta$ in 2SC/g2SC phase (dashed line)
and LOFF phase (solid line) and the LOFF momentum $q$ as functions
of $T$ for two coupling values $\eta=0.70$ and
$0.66$.\label{fig6}}
\end{center}
\end{figure}
\begin{figure}[!htb]
\begin{center}
\includegraphics[width=6cm]{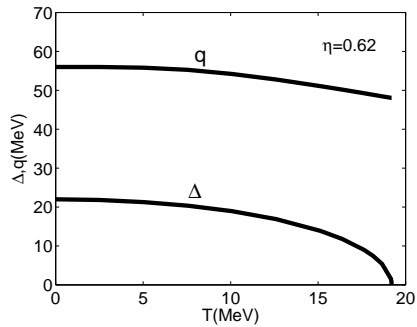}
\caption{The LOFF pairing gap $\Delta$ and LOFF momentum $q$ as
functions of $T$ at extremely weak coupling $\eta=0.63$ where the
2SC/g2SC phase starts to disappear. \label{fig7}}
\end{center}
\end{figure}

\subsection {Gluonic Phase}
The gluonic phase is energetically favored at strong coupling
where only the 4-7th gluons suffer instability and may appear at
intermediate coupling too\cite{gorbar2,gluonic}. In the gluonic
phase, off-diagonal gluon condensate or the spontaneously
generated off-diagonal Nambu-Goldstone current is nonzero. Since
the 4-7th gluons form a complex doublet, we can introduce only the
4-th gluon condensate
\begin{equation}
\vec{\rho}=g\langle\vec{A}_4\rangle=\langle\vec{\nabla}\varphi_4\rangle.
\end{equation}
Including this condensate, the thermodynamic potential of the
system becomes $\Omega(\Delta,\vec{\rho})$. Following the same
procedure used above, we can prove:
\\(1)The neutral gluonic phase with $\vec{\rho}\neq 0$ exists only in the temperature
region $T<T_4$ where the Meissner mass squared $m^2_4$ is
negative.
\\(2)The gluonic phase has lower free energy than the uniform phase
in the region $T<T_4$.
\\(3)At $T=T_4$, the value of $\vec{\rho}$ approaches to
zero continuously.

The proof is quite similar to that for the LOFF phase. We need
only the potential curvature definition of the Meissner masses
squared\cite{fukushima}
\begin{equation}
\partial^2 \Omega/\partial
\rho^2\big|_{\rho=0}=m_{4}^2/g^2
\end{equation}
and consider the fact that the gap equation for
$\vec\rho=(0,0,\rho)$ can be generally expressed as
\begin{equation}
\partial \Omega/\partial \rho=\rho K(\rho)=0.
\end{equation}
Recently, the numerical calculation on neutral gluonic phase is
presented at $T=0$ in \cite{gluonic}.

\section {Summary}
\label{s5}
We have investigated the stability of neutral two flavor color
superconducting quark matter at finite temperature. The main
conclusions are:
\\(1)There exists a temperature window below and close to the critical temperature of the superconductivity where the uniform 2SC/g2SC
phase is stable, namely the Meissner masses squared and the gap
susceptibility are both positive. The Meissner mass squared is
positive at temperatures $T_4<T<T_c$ for the 4-7th gluons and at
$T_8<T<T_c$ for the 8th gluon.
\\(2)In a wide range of coupling in the diquark channel, we have $T_4>T_8$, the introduction of LOFF phase can not completely solve the problem of
chromomagnetic instability at finite temperature, since in the
region $T_8<T<T_4$ only the 4-7th gluons suffer instability.
\\(3)The LOFF phase can exist only below the turning temperature $T_8$ where the Meissner mass squared for the 8th gluon changes
sign and the LOFF momentum approaches to zero. Similarly, the
gluonic phase can exist only below the turning temperature $T_4$
where the Meissner mass squared for the 4-7th gluons changes sign
and the off-diagonal gluon condensate vanishes.
\\(4)When the LOFF phase is taken into account, the strange temperature behavior of
the uniform color superconducting order parameter\cite{huang2}
disappears and the corresponding phase diagram of neutral quark
matter\cite{phase1,phase2,phase3,phase4,phase5} is significantly
changed. This situation is quite like the recent studies for
non-relativistic Fermi gas with population
inbalance\cite{chen,sedrakian2,he6}.

For further investigation, one needs to make a detailed
calculation for the phase with gluonic condensation or
Nambu-Goldstone current and to check the chromomagnetic stability
of the LOFF phase and gluonic phase. Since the LOFF momentum $\vec
q$ and the gluonic condensate $\vec\rho$ approach to zero
continuously at the turning temperatures $T_8$ and $T_4$, one can
at least expand the thermodynamic potential in terms of $\vec q$
and $\vec\rho$ in the neighborhood of $T_8$ and $T_4$. We defer
the research in this direction to be a future work.

{\bf Acknowledgement:} After having completed this work, we knew
that O.Kiriyama did a work\cite{kiriyama} where some results are
similar to ours. We thank him for useful discussions. This work is
supported by the grants NSFC10428510, 10435080 10575058 and
SRFDP20040003103.

\begin{widetext}
\appendix
\section{Functions $A, A', B, C, D, H$ and $J$}
\label{app}
We list in this appendix the functions $A, A', B, C, D, H$ and $J$
used to express the Meissner masses squared (\ref{meissner1}) and
(\ref{meissner2}) for the 4-8th gluons and photon. Using the
coefficients $C_{\pm\pm}^{ij}(p)$ and $C_{\pm\mp}^{ij}(p)$ defined
in \cite{huang3} and taking the trick of replacement
$p_0\rightarrow -p_0$ in calculating the Matsubara frequency
summation, we can prove that only 7 of the coefficients are
independent,
\begin{eqnarray}
&& C_{++}^{11}=C_{--}^{11}=A,\ \ \
C_{++}^{22}=C_{--}^{22}=A^\prime, \ \ \
C_{+-}^{12}=C_{+-}^{21}=C_{-+}^{12}=C_{-+}^{21}=B,\nonumber \\
&& C_{+-}^{11}=C_{-+}^{11}=C_{+-}^{22}=C_{-+}^{22}=C, \ \ \
C_{++}^{12}=C_{++}^{21}=C_{--}^{12}=C_{--}^{21}=D,\nonumber\\
&& C_{++}^{44}=C_{--}^{44}=H,\ \ \ C_{+-}^{44}=C_{-+}^{44}=J
\end{eqnarray}
with the explicit expressions of $A, A^\prime, B, C, D, H, J$ as
functions of $p\equiv |{\bf p}|$,
\begin{eqnarray}
A(p)&=&u_-^2v_-^2\frac{f(E_1)+f(E_2)-1}{E_\Delta^-}+u_+^2v_+^2\frac{f(E_3)+f(E_4)-1}{E_\Delta^+}+u_-^4f^\prime(E_2)+v_-^4f^\prime(E_1)
+u_+^4f^\prime(E_3)+v_+^4f^\prime(E_4), \nonumber\\
A^\prime(p)&=&u_-^2v_-^2\frac{f(E_1)+f(E_2)-1}{E_\Delta^-}+u_+^2v_+^2\frac{f(E_3)+f(E_4)-1}{E_\Delta^+}+u_-^4f^\prime(E_1)+v_-^4f^\prime(E_2)
+u_+^4f^\prime(E_4)+v_+^4f^\prime(E_3),\nonumber\\
B(p)&=&u_-^2v_-^2\frac{f(E_1)+f(E_2)-1}{E_\Delta^-}+u_+^2v_+^2\frac{f(E_3)+f(E_4)-1}{E_\Delta^+}-u_-^2v_-^2\left(f^\prime(E_1)+f^\prime(E_2)\right)
-u_+^2v_+^2\left(f^\prime(E_3)+f^\prime(E_4)\right),\nonumber\\
C(p)&=&\frac{(E_\Delta^-)^2-E_0^-E_0^+}{(E_\Delta^-)^2-(E_\Delta^+)^2}\frac{
f(E_1)+f(E_2)-1}{E_\Delta^-}-\frac{(E_\Delta^+)^2-E_0^-E_0^+}{(E_\Delta^-)^2-(E_\Delta^+)^2}\frac{
f(E_3)+f(E_4)-1}{E_\Delta^+}+\frac{1}{p},\nonumber\\
D(p)&=&-\frac{\Delta^2}{(E_\Delta^-)^2-(E_\Delta^+)^2}\left(\frac{f(E_1)+f(E_2)-1}{E_\Delta^-}-\frac{f(E_3)+f(E_4)-1}{E_\Delta^+}\right),\nonumber\\
H(p)&=&u_-^2\left(\frac{f(E_6)-f(E_2)}{E_6-E_2}+\frac{f(E_8)-f(E_1)}{E_8-E_1}\right)
+v_-^2\left(\frac{f(E_6)-f(-E_1)}{E_6+E_1}+\frac{f(E_8)-f(-E_2)}{E_8+E_2}\right)\nonumber\\
&+&u_+^2\left(\frac{f(E_5)-f(E_3)}{E_5-E_3}+\frac{f(E_7)-f(E_4)}{E_7-E_4}\right)
+v_+^2\left(\frac{f(E_5)-f(-E_4)}{E_5+E_4}+\frac{f(E_7)-f(-E_3)}{E_7+E_3}\right),\nonumber\\
J(p)&=&u_-^2\left(\frac{f(E_5)-f(-E_2)}{E_5+E_2}+\frac{f(E_7)-f(-E_1)}{E_7+E_1}\right)
+v_-^2\left(\frac{f(E_5)-f(E_1)}{E_5-E_1}+\frac{f(E_7)-f(E_2)}{E_7-E_2}\right),\nonumber\\
&+&u_+^2\left(\frac{f(E_6)-f(-E_3)}{E_6+E_3}+\frac{f(E_8)-f(-E_4)}{E_8+E_4}\right)
+v_+^2\left(\frac{f(E_6)-f(E_4)}{E_6-E_4}+\frac{f(E_8)-f(E_3)}{E_8-E_3}\right)+\frac{2}{p},
\end{eqnarray}
where the coherent coefficients $u_\pm^2$ and $v_\pm^2$ are
defined as $u_\pm^2=\left(1+E_0^\pm/E_\Delta^\pm\right)/2$ and
$v_\pm^2=\left(1-E_0^\pm/E_\Delta^\pm\right)/2$. Note that we have
added the terms $1/p$ to $C$ and $2/p$ to $J$ to cancel the vacuum
contribution. In this way the Meissner masses squared are
guaranteed to be zero in the normal phase with $\Delta=0$.
\end{widetext}

\end{document}